\documentclass[aps,twocolumn,superscriptaddress,preprintnumbers,pre]{revtex4}
\usepackage{xr-hyper}
\usepackage{mathrsfs, hyperref}
\usepackage{amssymb, amsbsy, amsmath, latexsym, dsfont, array, layout, graphics,mathrsfs,braket,amsfonts,amsthm,
  amssymb,graphicx,subfigure, youngtab,color,bm,mathtools,braket,verbatim,url,cleveref,natbib,hypernat,extarrows,inputenc,multirow,bbm}
  
\usepackage[usernames,dvipsnames]{xcolor}
\colorlet{RED}{red}
\usepackage{graphicx,psfrag}
\usepackage{amsfonts}
\usepackage[figuresright]{rotating}  
\usepackage{amssymb}
\usepackage{amsmath}
\usepackage{subfigure}
\usepackage{multirow}
\usepackage{tabularx}
\usepackage[resetlabels]{multibib}
\newcites{supp}{Supplemental References}
\usepackage{amssymb, amsbsy, amsmath, latexsym, dsfont, array, layout, graphics,mathrsfs,braket,amsfonts,amsthm,
  amssymb,graphicx,subfigure,dcolumn, youngtab,color,bm,mathtools,braket,verbatim,url,cleveref,natbib,hypernat,extarrows,inputenc,multirow}
\usepackage[usernames,dvipsnames]{xcolor}

\newcommand{\sss}[1]{{\color{Black} #1}}

\begin{document}

\title{Mirror-symmetry protected higher-order topological zero-frequency boundary and corner modes in Maxwell lattices} 

\author{Siddhartha Sarkar}
 \email{sarkarsi@umich.edu}
\author{Xiaoming Mao}
 \email{maox@umich.edu}
 \author{Kai Sun}
 \email{sunkai@umich.edu}
\affiliation{Department of Physics, University of Michigan, Ann Arbor, Michigan 48109, USA}

\date{\today}

\begin{abstract}
Maxwell lattices, where the number of degrees of freedom equals the number of constraints, are known to host topologically-protected zero-frequency modes and states of self stress, characterized by a topological index called topological polarization. In this letter, we show that in addition to these known topological modes, with the help of a mirror symmetry, the inherent chiral symmetry of Maxwell lattices creates another topological index, the mirror-graded winding number (MGWN). This MGWN is a higher order topological index, which gives rise to topological zero modes and states of self stress at mirror-invariant domain walls and corners between two systems with different MGWNs. We further show that two systems with same topological polarization can have different MGWNs, indicating that these two topological indices are fundamentally distinct. 

\end{abstract}
\maketitle
\textit{Introduction.}--Bulk-boundary correspondence is a defining feature of topological states where nontrivial topology of the bulk gives rise to modes localized at the boundary~\cite{HasanKane,QiZhang}. Early research on topological band theory focused on $d$-dimensional topological systems with localized states at $(d-1)$-dimensional boundaries (e.g., quantum Hall effect~\cite{KlitzingQHE}, quantum anomalous Hall effect~\cite{LiuZhangQiQAHE}, quantum spin Hall effect~\cite{KaneMeleQSHE,BHZQSHE}); this type of topology is now called first-order topology. 
A new kind of topological states, called higher-order topological states (HOTS), has been proposed in the last five years ~\cite{BenalcazarScience,BenalcazarPRB,schindler2018higher}. Here, instead of having $(d-1)$-dimensional topologically protected boundary modes, the $d$-dimensional $n$-th order topological system has $(d-n)$-dimensional ($n>1$) boundary modes. The boundary modes corresponding to $n=d$ and $n=d-1$ are generally called corner and hinge modes, respectively. These higher order states are generally protected by crystalline symmetries such as mirror~\cite{LangbehnReflection}, inversion~\cite{KhalafInversionHOTI}, rotation~\cite{SongRotationHOTI,GuidoOrtixRotoInversion}, product of time reversal (TRS) and rotation~\cite{schindler2018higher}, etc (see~\cite{xie2021higher} for an exhaustive literature survey). Along with  realizations in electronic systems, crystalline symmetry protected HOTS have been implemented in mechanical/elastic systems too, offering a class of materials in which elastic energy can be selectively confined to low-dimensional regions~\cite{PhysRevLett.122.204301,HatsugaiHOTI,serra2018observation,PhysRevResearch.1.032047,xue2019acoustic,ni2019observation}. 

One key challenge in the study of HOTS lies in the stability of topological corner modes. For example, in contrast to the quantum Hall effect, where the topological edge modes remain stable for {\it any} boundary conditions, for a 2D HOTS, unless certain special ingredient is introduced (e.g., a chiral symmetry), the frequency of the topological corner modes is in general not pinned to a particular value. Thus, depending on the microscopic details, such as boundary conditions and disorder near the corners, these topological modes 
and can disappear into bulk bands~\cite{BradlynCornerRobustness,OrtixCornerImmunity}. 
To overcome this challenge, recently, a generalized chiral symmetry was introduced to realize corner modes in an breathing kagome lattice acoustic metamaterial~\cite{ni2019observation}, while there are still some open discussions about the topological origin of these modes~\cite{OrtixCornerImmunity,PhysRevB.105.085411}. \sss{Another attempt~\cite{saremi2018controlling} showed existence of corner modes pinned at zero frequency in an over-constrained system made of rigid quadrilaterals connected by free hinges; however, this can be understood within the framework of boundary obstructed topological phases~\cite{khalaf2021boundary}.} 


In this Letter, we provide a different approach towards HOTS 
using Maxwell lattices (i.e., lattices with equal numbers of degrees of freedom (DOFs) $n_d$ and constraints $n_c$~\cite{maxwell1864calculation,lubensky2015phonons}), and show that the intrinsic chiral symmetry protected by this counting extends robustness to topological corner modes in this lattices, without requiring any detailed matching at boundaries. 
As shown by Kane and Lubensky~\cite{KaneLubensky}, Maxwell systems
can be mapped to a superconducting Bogoliubov de Gennes (BdG) Hamiltonian, which naturally has a chiral symmetry. With the BdG Hamiltonian, a first-order topological index, the topological polarization, can be introduced~\cite{KaneLubensky}, resulting in topologically protected edge modes at zero frequency.
We find that in addition to this first-order topological index, a nontrivial higher-order topological index (the MGWN~\cite{neupert2018topological,PhysRevLett.124.166804,imhof2018topolectrical}) can be introduced to a new class of Maxwell lattices, controlling zero-frequency topological domain-wall/corner modes, with robustness originating from the intrinsic chiral symmetry of the locking of degrees of freedom and constraints in Maxwell lattices.  
\noindent\textit{Kane-Lubensky topological index of Maxwell lattices.}--Linear mechanics of lattices made of point masses connected by springs is characterized by the compatibility matrix $\mathbf{C}$ which relates extensions of springs $e_i = C_{ij}u_j$ to the displacements $u_i$ of the point masses. Furthermore, $f_i = C^T_{ij}t_j$ relates the forces $f_i$ on the point masses to the tensions $t_i$ in the springs.  In Fourier space, the matrix $\mathbf{C}(\mathbf{q})$ has the size $n_c \times n_d$.  
The normal mode frequencies of these lattices $\omega^2(\mathbf{q})$ are the eigenvalues of the dynamical matrix $\mathbf{D}(\mathbf{q})= \mathbf{C}^\dagger (\mathbf{q}) \mathbf{C}(\mathbf{q})$ 
Kane and Lubensky~\cite{KaneLubensky} defined a `square root' of the dynamical matrix, which in reciprocal space takes the following form:
\begin{equation}
\mathcal{H}(\mathbf{q}) =\begin{pmatrix}
\mathbf{0} &  \mathbf{C}^\dagger(\mathbf{q})\\
 \mathbf{C}(\mathbf{q}) &\mathbf{0}
\end{pmatrix}.
\end{equation}
For every nonzero eigenvalue $\omega^2(\mathbf{q}) $ of $\mathbf{D}(\mathbf{q})$, $\mathcal{H}(\mathbf{q})$ has two eigenvalues $\pm \omega(\mathbf{q})$. The zero modes of $\mathcal{H}(\mathbf{q})$ include nullspace of $\mathbf{C}(\mathbf{q})$ (zero modes -- ZMs) and  nullspace of $\mathbf{C}^\dagger(\mathbf{q})$ (states of self stress -- SSSs), whereas the zero modes of $D(\mathbf{q})$ include the ZMs. Maxwell Calladine theorem~\cite{maxwell1864calculation,calladine1978buckminster} dictates that the number of ZMs ($n_0$) and number of SSSs ($n_s$) are equal ($n_0 = n_s$) for a Maxwell lattice. The matrix $\mathcal{H}(\mathbf{q})$ has the property that $S \mathcal{H}(\mathbf{q}) S = - \mathcal{H}(\mathbf{q})$, where $S =\text{Diag}\{\mathbbm{1},-\mathbbm{1}\}$. This property is known as the chiral (or sublattice) (anti)symmetry in the literature. Also, it is easy to check that $\mathcal{H}(\mathbf{q})$ has TRS: $\mathcal{H}(\mathbf{q}) = \mathcal{H}^*(-\mathbf{q})$, where $^*$ is complex conjugation. These two symmetries put the matrix $\mathcal{H}(\mathbf{q})$ in BDI class of Altland Zirnbauer classification~\cite{AltlandZirnbauer,kitaev2009periodic,ryu2010topological,RevModPhys.88.035005}. Along a closed loop $l$ in the Brillouin zone where the spectrum of the matrix is gapped at zero, a topological invariant $n_l$ can be defined:
$n_l = \frac{1}{2\pi i} \oint_l d\mathbf{q}\cdot \mathbf{\nabla}_\mathbf{q}\log \det \mathbf{C}^\dagger(\mathbf{q})$, which controls the number of topological ZMs at an open edge or  domain walls.

\noindent\textit{Mirror-graded winding number.}--Interestingly, in mirror symmetric Maxwell lattices, along the mirror invariant lines in the Brillouin zone, the mirror reflection operator $\mathbf{M}(\mathbf{q})$ commutes with the matrix $\mathcal{H}(\mathbf{q})$. Consequently, $\mathbf{M}(\mathbf{q})$ and $\mathcal{H}(\mathbf{q})$ can be simultaneously diagonalized. Since, $\mathbf{M}(\mathbf{q})$ only takes eigenvalues $\pm 1$, using the eigenvectors of $\mathbf{M}(\mathbf{q})$ the matrices $\mathbf{C}(\mathbf{q})$ and  $\mathcal{H}(\mathbf{q})$ can be block-diagonalized into odd $(-)$ and even $(+)$ sectors (Supplemental Material (SM)~\cite{SM2023} Sec.~\sss{SM.2-3}):
\begin{figure}[t]
\includegraphics{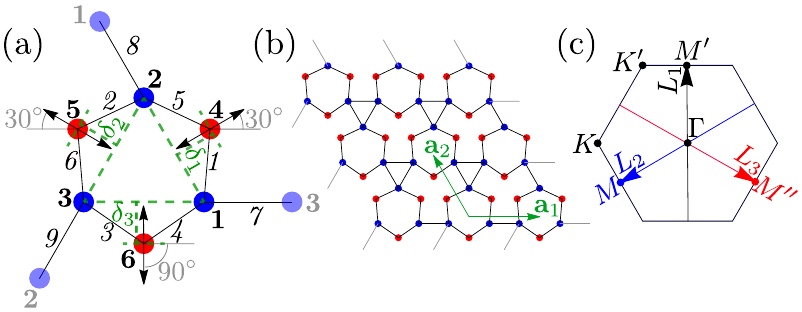}
\caption{The mirror symmetric Maxwell lattice. (a) The unit cell consists of three blue and three red point masses enumerated by bold numbers. The blue points can move in both $x$ and $y$ direction whereas the red points can only move along the direction of the corresponding double-directional black arrow. Parameter $\delta_i$ ($i = 1, \dots, 3$) is the perpendicular distance of point $i+3$ from the line joining points $i$ and $i+1$.  The numbers in italics enumerate the springs. The partially transparent blue points are in  adjacent unit cells. (b) A $3\times 3$ lattice. The springs shown in grey at the edges are required for periodic boundary condition. The green arrows show the the lattice vectors. (c) First Brillouin zone with the high symmetry points. The non-contractible loop $L_i$ is invariant under mirror reflection when $\delta_i = \delta_{i+1}$.} 
\label{Fig:figure1}
\end{figure}
\begin{subequations}
\label{Eq:BlockDiagonal}
\begin{align}
\mathbf{C}(\mathbf{q}) &= \begin{pmatrix}
\mathbf{C}_{-}(\mathbf{q}) & \mathbf{0}\\
 \mathbf{0} & \mathbf{C}_{+}(\mathbf{q})
\end{pmatrix},\\
\mathcal{H}(\mathbf{q}) &=\begin{pmatrix}
\mathbf{0} &  \mathbf{C}_{-}^\dagger(\mathbf{q}) & \mathbf{0} & \mathbf{0}\\
 \mathbf{C}_{-}(\mathbf{q}) & \mathbf{0} & \mathbf{0} & \mathbf{0}\\
\mathbf{0} & \mathbf{0} & \mathbf{0} &  \mathbf{C}_{+}^\dagger(\mathbf{q})\\
\mathbf{0} & \mathbf{0} &  \mathbf{C}_{+}(\mathbf{q}) & \mathbf{0}
\end{pmatrix}.
\end{align}
\end{subequations}
Now, using $ \mathbf{C}_{\pm}(\mathbf{q})$ we can define a topological invariant in each sector, the MGWNs:
\begin{equation}\label{Eq:MirrorWinding}
\nu_{\pm} = \frac{1}{2\pi i} \oint_{\mathbf{q}\rightarrow \mathbf{q}+\mathbf{G}_m} d\mathbf{q}\cdot \mathbf{\nabla}_\mathbf{q}\log \det \mathbf{C}_{\pm}^\dagger(\mathbf{q}),
\end{equation}
\begin{figure*}[t]
\includegraphics{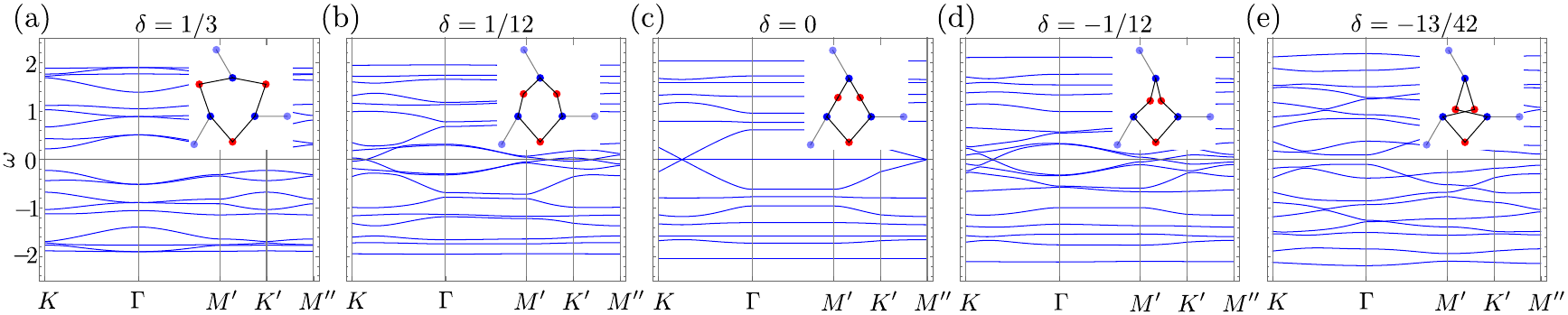}
\caption{Spectrum of $\mathcal{H}$ for different values $\delta = \delta_1 = \delta_2$ keeping $\delta_3 = 1/3$. The unit cell corresponding to each configuration is shown inset. Each diagram has $n_c+n_d = 18$ bands. All systems except (c) are gapped along line $\Gamma-M'$ ($L_1$ in Fig.~\ref{Fig:figure1}(c)). (c) has four flat bands at zero frequency. The spectrum in (b) and (d) are gapped at $\omega = 0$ along the line $\Gamma-M'$, but not gapped everywhere in the Brillouin zone. Only (a) and (e) are fully gapped at $\omega = 0$ over the entire Brillouin zone.} 
\label{Fig:figure2}
\end{figure*}
where $\mathbf{G}_m$ is the smallest reciprocal lattice vector along the mirror plane~\cite{neupert2018topological,PhysRevLett.124.166804,imhof2018topolectrical}. 
Note that $\nu_{+} + \nu_{-} = n_{l}$, since in this basis $\det \mathbf{C}(\mathbf{q}) = \det \mathbf{C}_{+}(\mathbf{q}) \det \mathbf{C}_{-}(\mathbf{q})$. In other words, the mirror symmetry allows us to split topological polarization into two different topological indices $\nu_{+}$ and $\nu_{-}$. This observation expanded the topological classification of Maxwell lattices, and allow us to realize HOTS.

It is worthwhile to highlight that to define a topological index, the Hamiltonian [Eq.~\eqref{Eq:BlockDiagonal}] must remain gapped with $\det \mathbf{C} \ne 0$. Because a mirror plane in the momentum space often passes through the $\Gamma$ point ($k=0$), it is necessary to gap the acoustic phonon bands at $\Gamma$. As will be shown below, this can be achieved by restricting the motion of certain lattice points, which break the translational invariance of the lattice. 


\noindent\textit{The mirror symmetric Maxwell lattice.}--We now illustrate one Maxwell lattice that support HOTS.
As shown in Fig.~\ref{Fig:figure1}, each unit cell of this lattice contains $6$ point masses with coordinates
\begin{subequations}
\begin{align}
\mathbf{r}_i&=\frac{1}{3} \left(\cos\left(\frac{2\pi i}{3}-\frac{5\pi}{6}\right),\sin\left(\frac{2\pi i}{3}-\frac{5\pi}{6}\right)\right),\\
\mathbf{r}_{i+3}&=(\frac{1}{6}+\delta_i) \left(\cos\left(\frac{2\pi i}{3}-\frac{\pi}{2}\right),\sin\left(\frac{2\pi i}{3}-\frac{\pi}{2}\right)\right),
\end{align}
\end{subequations}
with $i \in \{1,2,3\}$. The three points 
$i=1, 2$ and $3$ can move in both $x$ and $y$ directions, while the rest three are restricted to move along the direction marked by the black arrows shown in Fig.~\ref{Fig:figure1}(a):
\begin{subequations}
\begin{align}
\mathbf{u}_i &= \left(u_{ix},u_{iy}\right),\\
 \mathbf{u}_{i+3} &= u_{i+3}\left(\cos\left(\frac{2\pi i}{3}-\frac{\pi}{2}\right),\sin\left(\frac{2\pi i}{3}-\frac{\pi}{2}\right)\right),
\end{align}
\end{subequations}
for $i \in \{1,2,3\}$. Consequently, there are $n_d =9$ DOFs per unit cell $\{u_{1x},u_{1y},\dots,u_{3y},u_4,u_5,u_6\}$.

We then repeat this unit cell to form a 2D lattice and connect the mass poits with springs (solid lines in Fig.~\ref{Fig:figure1}(b)). Here we set the lattice vectors $\mathbf{a}_1 =(1,0)$ and $\mathbf{a}_2 =\frac{1}{2}(-1,\sqrt{3})$, and the masses of all points and the stiffnesses of all springs are set to 1 for simplicity. Notice that here we have $9$ springs per unit cell, which  match the DOFs $n_d =9$, making the system a Maxwell lattice.

Note that if we set $\delta_i = \delta_{i+1}$, the system is invariant under mirror reflection about the perpendicular bisector of points $i+3$ and $i+4$. The corresponding mirror invariant lines $L_i$ in the reciprocal space (Brillouin zone) are shown in Fig.~\ref{Fig:figure1}(c). Because all the mirror planes go through $\Gamma$, it is important to gap out the phonon bands at $\Gamma$ to define the topological index. In this setup, this is automatically achieved because points $i=4,5,6$ can only move along the arrow directions, which gaps out the acoustic modes.



\begin{figure}[b]
\includegraphics{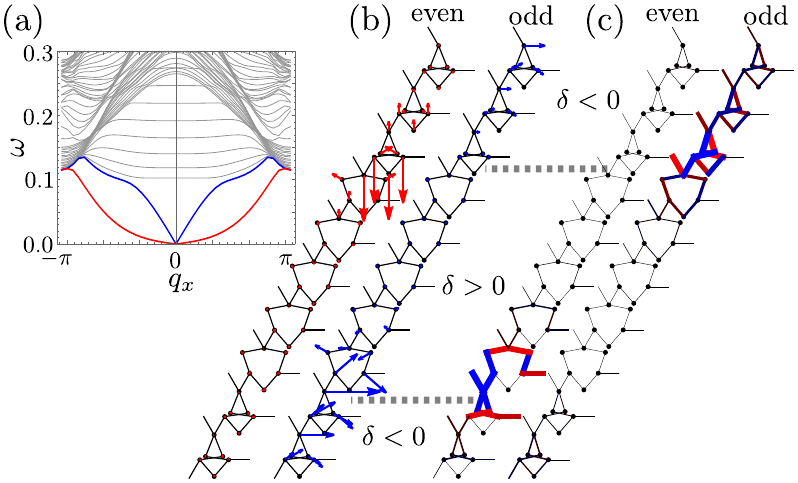}
\caption{Spectrum (a), ZMs (b) and SSSs (c) of a supercell consisting of $2N_0$ unit cells among which $N_0$ in the middle have $\delta = 1/3$ and the other ones have $\delta = -13/42$. Periodic boundary condition is employed in direction $(1/2,\sqrt{3}/2)$, whereas Bloch-periodic boundary condition $\mathbf{u}(\mathbf{x}+(1,0)) =\mathbf{u}(\mathbf{x}) e^{iq_x}$ is employed in $(1,0)$ direction. In (a), gray bands are bulk modes whereas the red and the blue bands are localized at the top and bottom domain walls, respectively. The left ZM in (b) is localized at the top domain wall and is even under vertical mirror $m_x$, whereas the right ZM in (b) is localized at the bottom domain wall and is odd under vertical mirror $m_x$. The left SSS in (c) is localized at the bottom domain wall and is even under vertical mirror $m_x$, whereas the right SSS in (c) is localized at the top domain wall and is odd under vertical mirror $m_x$. The red and blue colors of the bonds in (c) indicate the elongation and compression of the bonds, respectively.} 
\label{Fig:figure3}
\end{figure}

\begin{figure*}[t]
\includegraphics{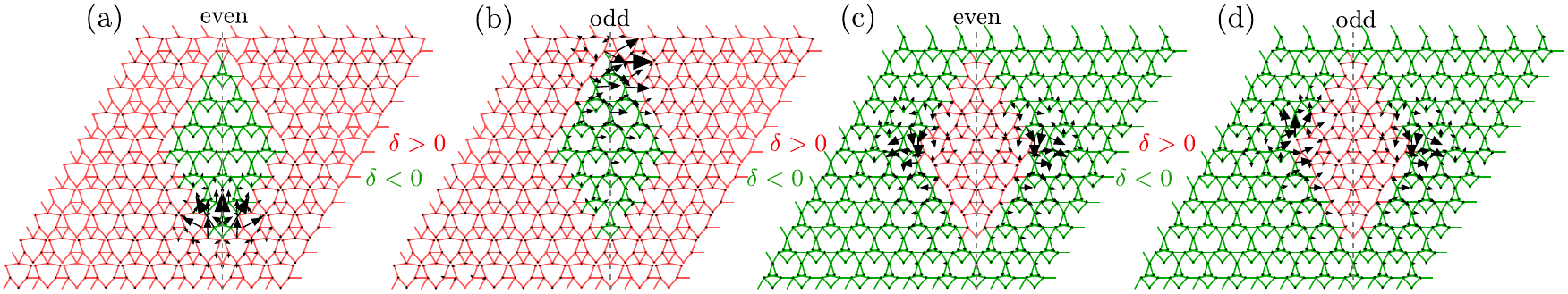}
\caption{Corner modes in systems with diamond shaped island of one phase inside the other phase. In each panel, the part of the system in red (green) has $\delta = 1/3$ ($\delta = -13/42$). The vertical grey dashed lines show the line of mirror symmetry; it passes through the the top and bottom corner of the diamond shaped island. We applied periodic boundary conditions in all cases. The black arrows show the displacement field corresponding to the zero modes. In all cases, the zero modes are concentrated at the corners. The corner modes in (a) and (c) are even under the vertical mirror reflection, whereas (b) and (d) are odd under the same reflection.
} 
\label{Fig:figure4}
\end{figure*}

The compatibility matrix $\mathbf{C}(\mathbf{q})$ is given in SM~\cite{SM2023} Sec.~\sss{SM.1}. For simplicity we set $\delta_3 = 1/3$ and vary $\delta_1 = \delta_2 \equiv \delta$. In this case, the lattice has one mirror $m_x$ per unit cell with normal in $x$ direction. Along the mirror invariant line $q_x = 0$ ($L_1$ in Fig.~\ref{Fig:figure1}(c)), we calculate $\det \left.\mathbf{C}_{+}^\dagger\right|_{L_1},\det \left.\mathbf{C}_{-}^\dagger\right|_{L_1}$ and integrate them 
from $\mathbf{q}=(0,-2\pi/\sqrt{3})$ to $\mathbf{q}+\mathbf{b}_2=(0,2\pi/\sqrt{3})$ along path $L_1$ according to Eq.~\eqref{Eq:MirrorWinding}.  We find
\begin{equation}
\nu_{+} = \begin{cases}
1 & \text{if } \delta >0,\\
0 & \text{if } \delta <0,
\end{cases}\;
\nu_{-} = \begin{cases}
0 & \text{if } \delta >0,\\
1 & \text{if } -5/12<\delta <0,\\
0 & \text{if } \delta <-5/12.
\end{cases}
\end{equation}
Clearly, the phases with $\delta > 0$ and $0>\delta>-5/12$ are distinct w.r.t. the MGWNs but same w.r.t. the Kane-Lubensky index. We will call $\delta>0$ phase 1, and $0>\delta>-5/12$ phase 2. In Fig.~\ref{Fig:figure2}, we show the spectrum of matrix $\mathcal{H}(\mathbf{q})$ for different values of $\delta$. At $\delta = 0$, the DOFs corresponding to points $\bold{4}$ and $\bold{5}$ are perpendicular to the springs connected to them; hence displacements of these points do not change the length of the springs to the linear order. These give two ZMs at every wave-vector $\mathbf{q}$. Then, due to the Maxwell-Calladine index theorem there are two SSSs at every  $\mathbf{q}$. Hence, there are $4$ flat bands at $\omega = 0$ of the matrix $\mathcal{H}(\mathbf{q})$ for $\delta = 0$. When $\delta \neq 0$, $\omega = 0$ gapped along $C_1$ line allowing us to define the MGWNs $\nu_{\pm}$.

In addition to defining the MGWNs, in order to localize ZMs
at the junction of two different mirror graded phases, we require the bulk bands to be completely gapped at $\omega = 0$ in addition to the path $L_1$. We find that phase 1 is fully gapped at $\omega = 0$ over the entire Brillouin zone for $\delta>5/42$ (Fig.~\ref{Fig:figure2}(a)), whereas phase 2 is fully gapped for $-5/12<\delta<-1/6$ (Fig.~\ref{Fig:figure2}(e)) (see SM~\cite{SM2023}~\sss{SM.4} for details).


\noindent\textit{Mirror-protected zero frequency edge states.}--To examine the bulk-edge correspondence, we create a supercell in Fig.~\ref{Fig:figure3} with periodic boundary conditions in both directions, which has domain walls separating $\delta =1/3$ and $\delta = -13/42$. The domain walls are horizontal -- normal to the mirror $m_x$; hence invariant under reflection about the mirror $m_x$. The spectrum of the dynamical matrix $\mathbf{D}(q_x)$ of the system is plotted as a function of surface wave vector $q_x$. We find two ZMs at $q_x = 0$ (Fig.~\ref{Fig:figure3}(a)). Since Kane-Lubensky indices of both domains are same: $n^{\delta = 1/3} = \nu^{\delta = 1/3}_{+} + \nu^{\delta = 1/3}_{-} = 1 = \nu^{\delta = -13/42}_{+} + \nu^{\delta = -13/42}_{-} =n^{\delta = -13/42}$, the ZMs at the domain walls are not given by the Kane-Lubensky index. However, since at $q_x = 0$, matrix $\mathbf{C}(\mathbf{q})$ can be block-diagonalized (Eq.~\eqref{Eq:BlockDiagonal}) as discussed above, we can use Eq.~\eqref{Eq:MirrorWinding} on $+$ and $-$ sectors separately. Since matrix $\mathbf{C}(q_x = 0, q_y)$ is block diagonal, the ZMs of each sector are also ZMs of the full system. Hence, at the top and bottom domain walls we get:
\begin{subequations}
\label{Eq:MirrorZMSSS}
\begin{align}
\text{top wall: } & \begin{split}
\nu_{+}^{<} - \nu_{+}^{>} &= 1 \Rightarrow \text{ ZM},\\
\nu_{-}^{<} - \nu_{-}^{>} &= -1 \Rightarrow \text{ SSS}, \end{split}\\
\text{bottom wall: } & \begin{split}
\nu_{+}^{<} - \nu_{+}^{>} &= -1 \Rightarrow \text{ SSS},\\
 \nu_{-}^{<} - \nu_{-}^{>} &= 1 \Rightarrow \text{ ZM},\end{split}
\end{align}
\end{subequations}
where $<$ and $>$ denote phases below and above the domain wall, respectively. It must be emphasized here that because rigid translation is not a zero mode in our lattice, in general such a lattice is not expected to have zero modes and all phonon modes should be gapped. However, at the domain boundary between regions with different topological indices, topological edge modes emerge with frequency pinned to zero by the chiral symmetry.

It is also worthwhile to highlight that these topological zero modes are fundamentally different from the zero modes protected by topological polarization. First of all, they are due to a totally different topological index. Secondly, in contrast to zero modes from topological polarization, the supercell spectrum of which has a flat bands at zero frequency~\cite{KaneLubensky}, the topological modes here are dispersive. Because the mirror symmetry is broken away from the mirror plane ($q_x\ne 0$), the frequency of the edge modes moves away from zero at $q_x\ne 0$ as shown in Fig.~\ref{Fig:figure3}. Finally, in contrast to the deformed kagome lattice (\cite{KaneLubensky}) where the SSSs and ZMs are localized on opposite domain walls, in our systems, the ZM and SSS are on the same domain wall. Typically, ZM and SSS cannot be localized on the same domain wall, because they will be lifted to finite frequency in the presence of hybridization between them. In our system, such hybridization is prohibited by the mirror symmetry, because for each domain, its ZM and SSS have opposite mirror parity (even vs odd).

To conclude this section, we would like to point out that this topological index and zero modes can also be characterized by a low-energy continuum theory (SM~\cite{SM2023} Sec.~\sss{SM.5}) using a Dirac Hamiltonian and the Jackiw-Rebbi analysis~\cite{JackiwRebbi,bernevig2013topological}.


\noindent\textit{Mirror-protected corner states.}--Mirror symmetric systems in the BDI class where the mirror reflection operator commutes TRS and chiral symmetry operators can have Mirror symmetry protected zero frequency corner modes~\cite{LangbehnReflection,GeierSeconOrder}. To look for such corner states, we create a diamond shaped island of one phase inside a rhombus shaped other phase phase with periodic boundary conditions for the rhombus in both direction (Fig.~\ref{Fig:figure4}). The top and the bottom corners of the diamond are invariant under a vertical mirror passing through them. In Figs.~\ref{Fig:figure4}(a-b), we see that when the inner island is $\delta = -13/42$ phase and the outer phase is $\delta = 1/3$, there are zero frequency corner modes localized at the top and the bottom corners, the top (bottom) one being odd (even) under the vertical mirror reflection. The situation is more curious when the inner island is $\delta = 1/3$ and the outer phase is $\delta = -13/42$ (Figs.~\ref{Fig:figure4}(c-d)). There are still two zero frequency corner modes, one of the odd and the other even under the vertical mirror reflection, but they are both localized at the right and left corners. 

The topological nature and the origin of these corner modes can be easily understood using standard approach of HOTS (SM~\cite{SM2023} Sec.~\sss{SM.6}). 
When the domain wall between the two phases is tilted such that the domain wall is not invariant under reflection, the localized states at the domain wall become massive, meaning that the spectrum is gapped at $\omega = 0$. Moreover, two oppositely tilted domain walls have opposite sign of the mass $m$; the sign of the mass $m$ depends on the sign of the angle of tilt of the domain wall. Therefore, at the corner both $\delta$ (across the domain boundary) and $m$ (along the domain boundary) change sign. 
As is elaborated in the SM~\cite{SM2023} Sec.~\sss{SM.6}, depending on the sign of the mass $m$, the amplitude of the zero frequency mode ($\sim e^{-m x}$) may either decrease or increases as we move away from the corner point ($x=0$). If the amplitude increases exponentially as we move away from this corner, it implies that this zero mode is localized at the next corner along the direction of the increasing amplitude. This theory analysis is in perfect agreement with numerical simulations. Furthermore, these corner modes persist even when the corner is not mirror invariant, as long as the bulk structures have mirror symmetry (see SM~\cite{SM2023} Sec.~\sss{SM.7}); which implies that this HOTS is ``intrinsic''~\cite{LangbehnReflection,GeierSeconOrder}.

\noindent\textit{Conclusions.}--In this work we demonstrated how spatial symmetries can protect higher order topological phase in Maxwell frames and gives rise to zero frequency topological edge and corner modes. Furthermore, these edge and corner modes are pinned to zero frequency due to inherent chiral symmetry of Maxwell frames pointed out in~\cite{KaneLubensky}. This chiral symmetry is often used as an approximate symmetry in fermionic systems (except in case of superconductors), but in case of Maxwell lattices it is exact. As mentioned earlier, our system falls under the BDI class of Altland-Zirnbauer classification; it has been known in the literature~\cite{LangbehnReflection,GeierSeconOrder} that mirror symmetry that commutes with time reversal and chiral symmetry can protect corner modes in 2-dimensions in this class. To our knowledge, our structure is the first example of this in classical systems. 
\sss{This system should be straightforwardly experimentally realized using hard plastic parts and hinges similar to what was done in~\cite{rocklin2017transformable} for deformed kagome lattice; with the three extra point masses (red points 4-6 in Fig.~\ref{Fig:figure1}(a)) in our system need to be put on fixed rails such that they can only move along the corresponding rails.}

\noindent \textit{Acknowledgements}.--S.S. thanks Xiaohan Wan for many discussions on this topic. This work was supported in part by the Office of Naval Research MURI N00014-20-1-2479.





\let\oldaddcontentsline\addcontentsline
\renewcommand{\addcontentsline}[3]{}
\bibliographystyle{apsrev4-1}
\bibliography{refer1}
\let\addcontentsline\oldaddcontentsline
\onecolumngrid
\vspace*{1cm}
\makeatletter
\renewcommand \thesection{S-\@arabic\c@section}
\renewcommand\thetable{S\@arabic\c@table}
\renewcommand \thefigure{S\@arabic\c@figure}
\renewcommand \theequation{S\@arabic\c@equation}
\makeatother
\setcounter{equation}{0}  
\setcounter{figure}{0}  
\setcounter{section}{0}  

{
    \center \bf \large 
    Supplemental Material\vspace*{0.1cm}\\ 
    \vspace*{0.0cm}
}
\maketitle
\tableofcontents
\let\oldsection\section
\renewcommand\section{\clearpage\oldsection}
\section{Compatibility matrix of the mirror symmetric Maxwell lattice}\label{Sec:Cmat}
The compatibility matrix corresponding to the system shown in Fig.~\sss{1} of the main text is given by:
\begin{equation}
\label{Eq:Cmat}
\tiny
\mathbf{C}(\mathbf{q}) = \begin{pmatrix}
\frac{1-6\delta_1}{2\sqrt{1+12\delta_1^2}} & \frac{-3-6\delta_1}{2\sqrt{3+36\delta_1^2}} & 0 & 0 & 0 & 0 & \frac{6\delta_1}{\sqrt{3+36\delta_1^2}} & 0 & 0\\
0 & 0 & \frac{1+6\delta_2}{2\sqrt{1+12\delta_2^2}} & \frac{3-6\delta_2}{2\sqrt{3+36\delta_2^2}} & 0 & 0 & 0 & \frac{6\delta_2}{\sqrt{3+36\delta_2^2}} & 0\\
0 & 0 & 0 & 0 & \frac{-1}{\sqrt{1+12\delta_3^2}} & \frac{6\delta_3}{\sqrt{3+36\delta_3^2}} & 0 & 0 & \frac{6\delta_3}{\sqrt{3+36\delta_3^2}}\\
 \frac{1}{\sqrt{1+12\delta_3^2}} & \frac{6\delta_3}{\sqrt{3+36\delta_3^2}} & 0 & 0 & 0 & 0 & 0 & 0 & \frac{6\delta_3}{\sqrt{3+36\delta_3^2}}\\
 0 & 0 &\frac{-1-6\delta_1}{2\sqrt{1+12\delta_1^2}} & \frac{3-6\delta_1}{2\sqrt{3+36\delta_1^2}} & 0 & 0 & \frac{6\delta_1}{\sqrt{3+36\delta_1^2}} & 0 & 0\\
 0 & 0 & 0 & 0 & \frac{-1+6\delta_2}{2\sqrt{1+12\delta_2^2}} & \frac{-3-6\delta_2}{2\sqrt{3+36\delta_2^2}} & 0 & \frac{6\delta_2}{\sqrt{3+36\delta_2^2}} & 0\\
 -1 & 0 & 0 & 0 & e^{i q_x} & 0 & 0 & 0 & 0\\
 -\frac{e^{i(-q_x+\sqrt{3} q_y)/2}}{2} & \frac{\sqrt{3}e^{i(-q_x+\sqrt{3} q_y)/2}}{2} & \frac{1}{2} & - \frac{\sqrt{3}}{2} & 0 & 0 & 0 & 0 & 0\\
 0 & 0 &  -\frac{e^{i(q_x+\sqrt{3} q_y)/2}}{2} & -\frac{\sqrt{3}e^{i(q_x+\sqrt{3} q_y)/2}}{2} & \frac{1}{2} & -\frac{\sqrt{3}}{2}& 0 & 0 & 0\\
\end{pmatrix}
\end{equation}

\section{Mirror symmetry, Block diagonalization of $\mathcal{H}$}\label{Sec:MirrorBlock}
When $\delta_1 = \delta_2  \equiv \delta$ in Fig.~\sss{1} of main text, the system is mirror symmetric about the vertical line passing through point $2$. Let us call this mirrior $m_x$ since its normal is in $x$-direction. Since $m_x$ flips the sign of the $x$ component of a vector, under this mirror the two lattice vectors (see Fig.~\sss{1} of main text) get mapped to
\begin{equation}
m_x \mathbf{a}_1 = -\mathbf{a}_1, m_x \mathbf{a}_2 = \mathbf{a}_1 + \mathbf{a}_2.
\end{equation}
As a consequence, a unit cell at $n_1 \mathbf{a}_1+ n_2 \mathbf{a}_2$ gets mapped to $(n_2-n_1)\mathbf{a}_1+n_2\mathbf{a}_2$. From Fig.~\sss{1} of main text, it is also easy to see that this mirror maps points $1\leftrightarrow 3$, $2\leftrightarrow 2$, $4\leftrightarrow 5$, $6\leftrightarrow 6$. With these information, we see that displacement states $|\mathbf{u}_i(n_1,n_2)\rangle$ transform under $m_x$ in the following way:
\begin{equation}
\begin{split}
m_x|\mathbf{u}_1(n_1,n_2)\rangle &= -\sigma_z |\mathbf{u}_3(n_2-n_1,n_2)\rangle,\\
m_x|\mathbf{u}_2(n_1,n_2)\rangle &= -\sigma_z |\mathbf{u}_2(n_2-n_1,n_2)\rangle,\\
m_x|\mathbf{u}_3(n_1,n_2)\rangle &= -\sigma_z |\mathbf{u}_1(n_2-n_1,n_2)\rangle,\\
m_x|u_4(n_1,n_2)\rangle &= |u_5(n_2-n_1,n_2)\rangle,\\
m_x|u_5(n_1,n_2)\rangle &= |u_4(n_2-n_1,n_2)\rangle,\\
m_x|u_6(n_1,n_2)\rangle &= |u_6(n_2-n_1,n_2)\rangle,
\end{split}
\end{equation}
where Pauli matrix $\sigma_z$ is used to flip the sign of the $y$ component of the vector, and we recall that the displacements of points 4, 5 and 6 are constrained. Defining the Fourier transforms of the displacement fields as $|\mathbf{u}_i(\mathbf{q})\rangle = \frac{1}{\sqrt{N}}\sum_{n_1,n_2}|\mathbf{u}_i(n_1,n_2)\rangle e ^{i \mathbf{q}\cdot(n_1\mathbf{a}_1+n_2\mathbf{a}_2)}$, we ask how these Fourier modes of displacements transform under the mirror. We show this below:
\begin{equation}
\begin{split}
m_x |\mathbf{u}_1(\mathbf{q})\rangle &=  \frac{1}{\sqrt{N}}\sum_{n_1,n_2}m_x|\mathbf{u}_1(n_1,n_2)\rangle e ^{i \mathbf{q}\cdot(n_1\mathbf{a}_1+n_2\mathbf{a}_2)}\\
&= \frac{1}{\sqrt{N}}\sum_{n_1,n_2}(-\sigma_z)|\mathbf{u}_3(n_2-n_1,n_2)\rangle e ^{i \mathbf{q}\cdot(n_1\mathbf{a}_1+n_2\mathbf{a}_2)}\\
&= \frac{1}{\sqrt{N}}\sum_{n_1',n_2'}(-\sigma_z)|\mathbf{u}_3(n_1',n_2')\rangle e ^{i \mathbf{q}\cdot((n_2'-n_1')\mathbf{a}_1+n_2'\mathbf{a}_2)}\\ 
&= \frac{1}{\sqrt{N}}\sum_{n_1',n_2'}(-\sigma_z)|\mathbf{u}_3(n_1',n_2')\rangle e ^{i \mathbf{q}\cdot((n_2'-n_1')(1,0)+n_2'(-1/2,\sqrt{3}/2))}\\
&= \frac{1}{\sqrt{N}}\sum_{n_1',n_2'}(-\sigma_z)|\mathbf{u}_3(n_1',n_2')\rangle e ^{i \mathbf{q}\cdot(n_2'/2-n_1',n_2'\sqrt{3}/2)}\\
&= \frac{1}{\sqrt{N}}\sum_{n_1',n_2'}(-\sigma_z)|\mathbf{u}_3(n_1',n_2')\rangle e ^{i (-q_x,q_y)\cdot(n_1'-n_2'/2,n_2'\sqrt{3}/2)}\\
&= \frac{1}{\sqrt{N}}\sum_{n_1',n_2'}(-\sigma_z)|\mathbf{u}_3(n_1',n_2')\rangle e ^{i (-q_x,q_y)\cdot(n_1'-n_2'/2,n_2'\sqrt{3}/2)}\\
&= \frac{1}{\sqrt{N}}\sum_{n_1',n_2'}(-\sigma_z)|\mathbf{u}_3(n_1',n_2')\rangle e ^{i (-q_x,q_y)\cdot(n_1'\mathbf{a}_1+n_2'\mathbf{a}_2)}\\
&= (-\sigma_z)|\mathbf{u}_3(-q_x,q_y)\rangle\\
&= (-\sigma_z)|\mathbf{u}_3(m_x\mathbf{q})\rangle.
\end{split}
\end{equation}
Similarly,
\begin{equation}
\begin{split}
m_x |\mathbf{u}_2(\mathbf{q})\rangle &= (-\sigma_z)|\mathbf{u}_2(m_x\mathbf{q})\rangle\\
m_x |\mathbf{u}_3(\mathbf{q})\rangle &= (-\sigma_z)|\mathbf{u}_1(m_x\mathbf{q})\rangle\\
m_x |u_4(\mathbf{q})\rangle &=|u_5(m_x\mathbf{q})\rangle\\
m_x |u_5(\mathbf{q})\rangle &= |u_4(m_x\mathbf{q})\rangle\\
m_x |u_6(\mathbf{q})\rangle &= |u_6(m_x\mathbf{q})\rangle
\end{split}
\end{equation}
All together, the transformation is the following:
\begin{equation}
\begin{split}
&m_x\{|u_{1x}(\mathbf{q})\rangle |u_{1y}(\mathbf{q})\rangle, |u_{2x}(\mathbf{q})\rangle, |u_{2y}(\mathbf{q})\rangle, |u_{3x}(\mathbf{q})\rangle, |u_{3y}(\mathbf{q})\rangle, |u_{4}(\mathbf{q})\rangle, |u_{5}(\mathbf{q})\rangle, |u_{6}(\mathbf{q})\rangle\} \\
=& \{|u_{1x}(m_x\mathbf{q})\rangle, |u_{1y}(m_x\mathbf{q})\rangle, |u_{2x}(m_x\mathbf{q})\rangle, |u_{2y}(m_x\mathbf{q})\rangle, |u_{3x}(m_x\mathbf{q})\rangle, |u_{3y}(m_x\mathbf{q})\rangle, |u_{4}(m_x\mathbf{q})\rangle, |u_{5}(m_x\mathbf{q})\rangle, |u_{6}(m_x\mathbf{q})\rangle\}\mathbf{M}_u(m_x\mathbf{q}),
\end{split}
\end{equation}
where
\begin{equation}
\mathbf{M}_u(\mathbf{q}) =\begin{pmatrix}
0 & 0 & 0 & 0 & -1 & 0 & 0 & 0 & 0\\
0 & 0 & 0 & 0 & 0 & 1 & 0 & 0 & 0\\
0 & 0 & -1 & 0 & 0 & 0 & 0 & 0 & 0\\
0 & 0 & 0 & 1 & 0 & 0 & 0 & 0 & 0\\
-1 & 0 & 0 & 0 & 0 & 0 & 0 & 0 & 0\\
0 & 1 & 0 & 0 & 0 & 0 & 0 & 0 & 0\\
0 & 0 & 0 & 0 & 0 & 0 & 0 & 1 & 0\\
0 & 0 & 0 & 0 & 0 & 0 & 1 & 0 & 0\\
0 & 0 & 0 & 0 & 0 & 0 & 0 & 0 & 1
\end{pmatrix}.
\end{equation}
Now, we turn to the bonds. Under mirror, the bond elongation states get mapped the following way:
\begin{equation}
\begin{split}
m_x|e_1(n_1,n_2)\rangle &= |e_6(n_2-n_1,n_2)\rangle,\\
m_x|e_2(n_1,n_2)\rangle &= |e_5(n_2-n_1,n_2)\rangle,\\
m_x|e_3(n_1,n_2)\rangle &= |e_4(n_2-n_1,n_2)\rangle,\\
m_x|e_4(n_1,n_2)\rangle &= |e_3(n_2-n_1,n_2)\rangle,\\
m_x|e_5(n_1,n_2)\rangle &= |e_2(n_2-n_1,n_2)\rangle,\\
m_x|e_6(n_1,n_2)\rangle &= |e_1(n_2-n_1,n_2)\rangle,\\
m_x|e_7(n_1,n_2)\rangle &= |e_7(n_2-n_1-1,n_2)\rangle,\\
m_x|e_8(n_1,n_2)\rangle &= |e_9(n_2-n_1+1,n_2+1)\rangle,\\
m_x|e_9(n_1,n_2)\rangle &= |e_8(n_2-n_1,n_2-1)\rangle.
\end{split}
\end{equation}
Note that the transformation of the last three bonds are different because they are inter-unit-cell bonds.  Define the Fourier transforms of the bond elongation states as $|e_i(\mathbf{q})\rangle = \frac{1}{\sqrt{N}}\sum_{n_1,n_2}|e_i(n_1,n_2)\rangle e ^{i \mathbf{q}\cdot(n_1\mathbf{a}_1+n_2\mathbf{a}_2)}$. The transformation of Fourier modes of the first 6 bonds under $m_x$ can be obtained similar to the displacements:
\begin{equation}
\begin{split}
m_x |e_1(\mathbf{q})\rangle &= |e_6(m_x\mathbf{q})\rangle,\\
m_x |e_2(\mathbf{q})\rangle &= |e_5(m_x\mathbf{q})\rangle,\\
m_x |e_3(\mathbf{q})\rangle &= |e_4(m_x\mathbf{q})\rangle,\\
m_x |e_4(\mathbf{q})\rangle &= |e_3(m_x\mathbf{q})\rangle,\\
m_x |e_5(\mathbf{q})\rangle &= |e_2(m_x\mathbf{q})\rangle,\\
m_x |e_6(\mathbf{q})\rangle &= |e_1(m_x\mathbf{q})\rangle.
\end{split}
\end{equation}
The transformation of the Fourier mode of the 7th bond is as follows:
\begin{equation}
\begin{split}
m_x |e_7(\mathbf{q})\rangle &=  \frac{1}{\sqrt{N}}\sum_{n_1,n_2}m_x|e_7(n_1,n_2)\rangle e ^{i \mathbf{q}\cdot(n_1\mathbf{a}_1+n_2\mathbf{a}_2)}\\
&= \frac{1}{\sqrt{N}}\sum_{n_1,n_2}|e_7(n_2-n_1-1,n_2)\rangle e ^{i \mathbf{q}\cdot(n_1\mathbf{a}_1+n_2\mathbf{a}_2)}\\
&= \frac{1}{\sqrt{N}}\sum_{n_1',n_2'}|e_7(n_1',n_2')\rangle e ^{i \mathbf{q}\cdot((n_2'-n_1'-1)\mathbf{a}_1+n_2'\mathbf{a}_2)}\\ 
&= \frac{1}{\sqrt{N}}\sum_{n_1',n_2'}|e_7(n_1',n_2')\rangle e ^{i \mathbf{q}\cdot((n_2'-n_1'-1)(1,0)+n_2'(-1/2,\sqrt{3}/2))}\\
&= \frac{1}{\sqrt{N}}\sum_{n_1',n_2'}|e_7(n_1',n_2')\rangle e ^{i \mathbf{q}\cdot(n_2'/2-n_1'-1,n_2'\sqrt{3}/2)}\\
&= \frac{1}{\sqrt{N}}\sum_{n_1',n_2'}|e_7(n_1',n_2')\rangle e ^{i (-q_x,q_y)\cdot(n_1'-n_2'/2,n_2'\sqrt{3}/2)} e ^{-i q_x}\\
&= \frac{1}{\sqrt{N}}\sum_{n_1',n_2'}|e_7(n_1',n_2')\rangle e ^{i (-q_x,q_y)\cdot(n_1'-n_2'/2,n_2'\sqrt{3}/2)}e ^{-i q_x}\\
&= \frac{1}{\sqrt{N}}\sum_{n_1',n_2'}|e_7(n_1',n_2')\rangle e ^{i (-q_x,q_y)\cdot(n_1'\mathbf{a}_1+n_2'\mathbf{a}_2)}e ^{-i q_x}\\
&=e ^{-i q_x}|e_7(-q_x,q_y)\rangle\\
&= e ^{-i q_x}|e_7(m_x\mathbf{q})\rangle.
\end{split}
\end{equation}
Similarly, for bond 8
\begin{equation}
\begin{split}
m_x |e_8(\mathbf{q})\rangle &=  \frac{1}{\sqrt{N}}\sum_{n_1,n_2}m_x|e_8(n_1,n_2)\rangle e ^{i \mathbf{q}\cdot(n_1\mathbf{a}_1+n_2\mathbf{a}_2)}\\
&= \frac{1}{\sqrt{N}}\sum_{n_1,n_2}|e_9(n_2-n_1+1,n_2+1)\rangle e ^{i \mathbf{q}\cdot(n_1\mathbf{a}_1+n_2\mathbf{a}_2)}\\
&= \frac{1}{\sqrt{N}}\sum_{n_1',n_2'}|e_9(n_1',n_2')\rangle e ^{i \mathbf{q}\cdot((n_2'-n_1')\mathbf{a}_1+(n_2'-1)\mathbf{a}_2)}\\ 
&= \frac{1}{\sqrt{N}}\sum_{n_1',n_2'}|e_9(n_1',n_2')\rangle e ^{i \mathbf{q}\cdot((n_2'-n_1')(1,0)+(n_2'-1)(-1/2,\sqrt{3}/2))}\\
&= \frac{1}{\sqrt{N}}\sum_{n_1',n_2'}|e_9(n_1',n_2')\rangle e ^{i \mathbf{q}\cdot(n_2'/2-n_1',n_2'\sqrt{3}/2)}e^{-i(-q_x+\sqrt{3}q_y)/2}\\
&= \frac{1}{\sqrt{N}}\sum_{n_1',n_2'}|e_9(n_1',n_2')\rangle e ^{i (-q_x,q_y)\cdot(n_1'-n_2'/2,n_2'\sqrt{3}/2)} e^{-i(-q_x+\sqrt{3}q_y)/2}\\
&= \frac{1}{\sqrt{N}}\sum_{n_1',n_2'}|e_9(n_1',n_2')\rangle e ^{i (-q_x,q_y)\cdot(n_1'-n_2'/2,n_2'\sqrt{3}/2)}e^{-i(-q_x+\sqrt{3}q_y)/2}\\
&= \frac{1}{\sqrt{N}}\sum_{n_1',n_2'}|e_9(n_1',n_2')\rangle e ^{i (-q_x,q_y)\cdot(n_1'\mathbf{a}_1+n_2'\mathbf{a}_2)}e^{-i(-q_x+\sqrt{3}q_y)/2}\\
&=e^{-i(-q_x+\sqrt{3}q_y)/2}|e_9(-q_x,q_y)\rangle\\
&= e^{-i(-q_x+\sqrt{3}q_y)/2}|e_9(m_x\mathbf{q})\rangle,
\end{split}
\end{equation}
and for bond 9
\begin{equation}
\begin{split}
m_x |e_9(\mathbf{q})\rangle &=  \frac{1}{\sqrt{N}}\sum_{n_1,n_2}m_x|e_9(n_1,n_2)\rangle e ^{i \mathbf{q}\cdot(n_1\mathbf{a}_1+n_2\mathbf{a}_2)}\\
&= \frac{1}{\sqrt{N}}\sum_{n_1,n_2}|e_8(n_2-n_1,n_2-1)\rangle e ^{i \mathbf{q}\cdot(n_1\mathbf{a}_1+n_2\mathbf{a}_2)}\\
&= \frac{1}{\sqrt{N}}\sum_{n_1',n_2'}|e_8(n_1',n_2')\rangle e ^{i \mathbf{q}\cdot((n_2'+1-n_1')\mathbf{a}_1+(n_2'+1)\mathbf{a}_2)}\\ 
&= \frac{1}{\sqrt{N}}\sum_{n_1',n_2'}|e_8(n_1',n_2')\rangle e ^{i \mathbf{q}\cdot((n_2'+1-n_1')(1,0)+(n_2'+1)(-1/2,\sqrt{3}/2))}\\
&= \frac{1}{\sqrt{N}}\sum_{n_1',n_2'}|e_8(n_1',n_2')\rangle e ^{i \mathbf{q}\cdot(n_2'/2-n_1',n_2'\sqrt{3}/2)}e^{-i(-q_x-\sqrt{3}q_y)/2}\\
&= \frac{1}{\sqrt{N}}\sum_{n_1',n_2'}|e_8(n_1',n_2')\rangle e ^{i (-q_x,q_y)\cdot(n_1'-n_2'/2,n_2'\sqrt{3}/2)} e^{-i(-q_x-\sqrt{3}q_y)/2}\\
&= \frac{1}{\sqrt{N}}\sum_{n_1',n_2'}|e_8(n_1',n_2')\rangle e ^{i (-q_x,q_y)\cdot(n_1'-n_2'/2,n_2'\sqrt{3}/2)}e^{-i(-q_x-\sqrt{3}q_y)/2}\\
&= \frac{1}{\sqrt{N}}\sum_{n_1',n_2'}|e_8(n_1',n_2')\rangle e ^{i (-q_x,q_y)\cdot(n_1'\mathbf{a}_1+n_2'\mathbf{a}_2)}e^{-i(-q_x-\sqrt{3}q_y)/2}\\
&=e^{-i(-q_x-\sqrt{3}q_y)/2}|e_8(-q_x,q_y)\rangle\\
&= e^{-i(-q_x-\sqrt{3}q_y)/2}|e_8(m_x\mathbf{q})\rangle.
\end{split}
\end{equation}
All together, the transformation is the following:
\begin{equation}
\begin{split}
&m_x\{|e_{1}(\mathbf{q})\rangle |e_{2}(\mathbf{q})\rangle, |e_{3}(\mathbf{q})\rangle, |e_{4}(\mathbf{q})\rangle, |e_{5}(\mathbf{q})\rangle, |e_{6}(\mathbf{q})\rangle, |e_{7}(\mathbf{q})\rangle, |e_{8}(\mathbf{q})\rangle, |e_{9}(\mathbf{q})\rangle\} \\
=& \{|e_{1}(m_x\mathbf{q})\rangle, |e_{2}(m_x\mathbf{q})\rangle, |e_{3}(m_x\mathbf{q})\rangle, |e_{4}(m_x\mathbf{q})\rangle, |e_{5}(m_x\mathbf{q})\rangle, |e_{6}(m_x\mathbf{q})\rangle, |e_{7}(m_x\mathbf{q})\rangle, |e_{8}(m_x\mathbf{q})\rangle, |e_{9}(m_x\mathbf{q})\rangle\}\mathbf{M}_e(m_x\mathbf{q}),
\end{split}
\end{equation}
where,
\begin{equation}
\mathbf{M}_e(\mathbf{q}) =\begin{pmatrix}
0 & 0 & 0 & 0 & 0 & 1 & 0 & 0 & 0\\
0 & 0 & 0 & 0 & 1 & 0 & 0 & 0 & 0\\
0 & 0 & 0 & 1 & 0 & 0 & 0 & 0 & 0\\
0 & 0 & 1 & 0 & 0 & 0 & 0 & 0 & 0\\
0 & 1 & 0 & 0 & 0 & 0 & 0 & 0 & 0\\
1 & 0 & 0 & 0 & 0 & 0 & 0 & 0 & 0\\
0 & 0 & 0 & 0 & 0 & 0 & e ^{i q_x} & 0 & 0\\
0 & 0 & 0 & 0 & 0 & 0 & 0 & 0 & e^{-i(q_x-\sqrt{3}q_y)/2}\\
0 & 0 & 0 & 0 & 0 & 0 & 0 & e^{-i(q_x+\sqrt{3}q_y)/2} & 0
\end{pmatrix}.
\end{equation}
With these, now we are at a position to find how the compatibility matrix transforms under $m_x$. As an operator that act on the displacement space to give the elongations of the bonds, the compatibility operator can be written as:
\begin{equation}
\hat{\mathbf{C}} =  \sum_{\mathbf{R},\mathbf{R}'}  \sum_{i,j} |e_i(\mathbf{R})\rangle C_{ij}(\mathbf{R}-\mathbf{R}')\langle u_j(\mathbf{R}')|,
\end{equation}
where $\mathbf{R}$ and $\mathbf{R}'$ are positions of the unit cells, and $i$ goes over all $9$ the 9 bonds in each unit cell whereas $j$ goes over the 9 degrees of freedom in each unit cell. We can write this in terms of the Fourier modes in the following way:
\begin{equation}
\begin{split}
\hat{\mathbf{C}} &=  \sum_{\mathbf{R},\mathbf{R}'}  \sum_{i,j} |e_i(\mathbf{R})\rangle C_{ij}(\mathbf{R}-\mathbf{R}')\langle u_j(\mathbf{R}') |\\
&=  \sum_{\mathbf{R},\mathbf{R}'}  \sum_{i,j}  \frac{1}{N}\sum_{\mathbf{q},\mathbf{q}'} |e_i(\mathbf{q})\rangle e^{-i \mathbf{q}\cdot \mathbf{R}} C_{ij}(\mathbf{R}-\mathbf{R}') e^{i \mathbf{q}'\cdot \mathbf{R}'}\langle u_j(\mathbf{q}') |\\
&=  \sum_{\mathbf{q},\mathbf{q}'}\sum_{i,j} |e_i(\mathbf{q})\rangle \langle u_j(\mathbf{q}') |  \sum_{\mathbf{R},\mathbf{R}'}  \frac{1}{N} e^{-i \mathbf{q}\cdot \mathbf{R}} C_{ij}(\mathbf{R}-\mathbf{R}') e^{i \mathbf{q}'\cdot \mathbf{R}'}\\
&=  \sum_{\mathbf{q},\mathbf{q}'}\sum_{i,j} |e_i(\mathbf{q})\rangle \langle u_j(\mathbf{q}') |  \sum_{\mathbf{R},\mathbf{R}''}  \frac{1}{N} e^{-i \mathbf{q}\cdot \mathbf{R}} C_{ij}(\mathbf{R}'') e^{i \mathbf{q}'\cdot (\mathbf{R}-\mathbf{R}'')}\\
&=  \sum_{\mathbf{q},\mathbf{q}'}\sum_{i,j} |e_i(\mathbf{q})\rangle \langle u_j(\mathbf{q}') |  \sum_{\mathbf{R}''}  \frac{1}{N} C_{ij}(\mathbf{R}'') e^{i \mathbf{q}'\cdot (-\mathbf{R}'')} N \delta_{\mathbf{q}-\mathbf{q}',,\mathbf{0}}\\
&=  \sum_{\mathbf{q}} \sum_{i,j}|e_i(\mathbf{q})\rangle \langle u_j(\mathbf{q}) |  \sum_{\mathbf{R}''} C_{ij}(\mathbf{R}'') e^{-i \mathbf{q}\cdot \mathbf{R}''}\\
&=  \sum_{\mathbf{q}} \sum_{i,j} |e_i(\mathbf{q})\rangle  C_{ij}(\mathbf{q}) \langle u_j(\mathbf{q}) |,
\end{split}
\end{equation}
where we used the definition  $C_{ij}(\mathbf{q}) = \sum_{\mathbf{R}} C_{ij}(\mathbf{R}) e^{-i \mathbf{q}\cdot \mathbf{R}}$ and the identity $\sum_{\mathbf{R}}e^{i\mathbf{q}\cdot \mathbf{R}} = N \delta_{\mathbf{q},\mathbf{0}}$, where $\delta_{i,j}$ is the Kronecker delta function. We understand that $C_{ij}(\mathbf{q})$ are the elements of the matrix $\mathbf{C}(\mathbf{q})$ in Eq.~\eqref{Eq:Cmat}. Since, the system is invariant under the mirror $m_x$, the operator $\hat{\mathbf{C}}$ is also invariant under $m_x$. This has the following consequence:
\begin{equation}
\begin{split}
\hat{\mathbf{C}} &= m_x \hat{\mathbf{C}}  m_x^{\dagger},\\
\Rightarrow \sum_{\mathbf{q}} \sum_{i,j} |e_i(\mathbf{q})\rangle  C_{ij}(\mathbf{q}) \langle u_j(\mathbf{q})| &= \sum_{\mathbf{q}} \sum_{i,j}  m_x |e_i(\mathbf{q})\rangle  C_{ij}(\mathbf{q}) \langle u_j(\mathbf{q})|   m_x^{\dagger}\\
&= \sum_{\mathbf{q}} \sum_{i,j} \sum_{i',j'}  |e_{i'}(m_x\mathbf{q})\rangle M_ e(m_x\mathbf{q})_{i'i} C_{ij}(\mathbf{q}) M_ u(m_x\mathbf{q})_{j'j}^* \langle u_{j'}(m_x\mathbf{q})| \\
&= \sum_{q_x,q_y} \sum_{i,j} \sum_{i',j'}  |e_{i'}(-q_x,q_y)\rangle M_ e(-q_x,q_y)_{i'i} C_{ij}(q_x,q_y) M_ u(-q_x,q_y)_{j'j}^* \langle u_{j'}(-q_x,q_y)| \\ 
&= \sum_{q_x,q_y} \sum_{i,j} \sum_{i',j'}  |e_i(q_x,q_y)\rangle M_ e(q_x,q_y)_{ii'} C_{i'j'}(-q_x,q_y) M_ u(q_x,q_y)_{jj'}^* \langle u_j(-q_x,q_y)|, \\
\Rightarrow  C_{ij}(\mathbf{q}) &= \sum_{i',j'} M_ e(q_x,q_y)_{ii'} C_{i'j'}(-q_x,q_y) M_ u(q_x,q_y)_{jj'}^*,\\
\Rightarrow  \mathbf{C}(\mathbf{q}) &= \mathbf{M}_ e(\mathbf{q}) \mathbf{C}(m_x\mathbf{q}) \mathbf{M}_ u(\mathbf{q})^\dagger ,\\
\Rightarrow  \mathbf{M}_ e^\dagger(\mathbf{q})  \mathbf{C}(\mathbf{q})\mathbf{M}_ u(\mathbf{q}) &= \mathbf{C}(m_x\mathbf{q}).
\end{split}
\end{equation}
Then, the ``square root" Hamiltonian $\mathcal{H}(\mathbf{q})$ transforms as the following:
\begin{equation}
\begin{pmatrix}
\mathbf{M}_u^\dagger(\mathbf{q}) & \mathbf{0}\\
\mathbf{0} & \mathbf{M}_e^\dagger(\mathbf{q})
\end{pmatrix} 
\begin{pmatrix}
\mathbf{0} & \mathbf{C}^\dagger(\mathbf{q})\\
 \mathbf{C}(\mathbf{q}) & \mathbf{0}
\end{pmatrix}  
\begin{pmatrix}
\mathbf{M}_u(\mathbf{q}) & \mathbf{0}\\
\mathbf{0} & \mathbf{M}_e(\mathbf{q})
\end{pmatrix}
= 
\begin{pmatrix}
\mathbf{0} & \mathbf{C}^\dagger(m_x\mathbf{q})\\
 \mathbf{C}(m_x \mathbf{q}) & \mathbf{0}
\end{pmatrix} \Rightarrow \mathbf{M}^\dagger(\mathbf{q}) \mathcal{H}(\mathbf{q}) \mathbf{M}(\mathbf{q}) = \mathcal{H}(m_x\mathbf{q}),
\end{equation}
where
\begin{equation}
\mathbf{M}(\mathbf{q}) = \begin{pmatrix}
\mathbf{M}_u(\mathbf{q}) & \mathbf{0}\\
\mathbf{0} & \mathbf{M}_e(\mathbf{q})
\end{pmatrix}.
\end{equation}
Since $m_x^2 = \mathbbm{1}$, the matrices $\mathbf{M}_u(\mathbf{q})$, $\mathbf{M}_e(\mathbf{q})$ and $\mathbf{M}(\mathbf{q})$ have the following property:
\begin{equation}
\mathbf{M}_u(m_x\mathbf{q})\mathbf{M}_u(\mathbf{q})=\mathbbm{1},\mathbf{M}_e(m_x\mathbf{q})\mathbf{M}_e(\mathbf{q})=\mathbbm{1},\mathbf{M}(m_x\mathbf{q})\mathbf{M}(\mathbf{q})=\mathbbm{1}.
\end{equation}
Therefore, on the line where $m_x\mathbf{q} = \mathbf{q} \Rightarrow q_x = 0$, the following is true: $\mathbf{M}_u^2(q_x = 0,q_y) = \mathbbm{1}$, $\mathbf{M}_e^2(q_x = 0,q_y)  = \mathbbm{1}$, $\mathbf{M}^2(q_x = 0,q_y)  = \mathbbm{1}$. Hence, on the line $q_x = 0$, the eigenvalues of $\mathbf{M}_u$, $\mathbf{M}_e$ and $\mathbf{M}$ are $\pm 1$. The eigenvectors of $\mathbf{M}_u(q_x = 0,q_y)$ and $\mathbf{M}_e(q_x = 0,q_y)$ are listed below
\begin{equation}
\begin{split}
e_{1-}^{(u)} = \frac{1}{\sqrt{2}}\begin{bmatrix}
0\\ 0\\ 0\\ 0\\ 0\\ 0\\ -1\\ 1\\ 0
\end{bmatrix}, e_{2-}^{(u)} = \frac{1}{\sqrt{2}}\begin{bmatrix}
0\\ -1\\ 0\\ 0\\ 0\\ 1\\ 0\\ 0\\ 0
\end{bmatrix}, e_{3-}^{(u)} = \frac{1}{\sqrt{2}}\begin{bmatrix}
1\\ 0\\ 0\\ 0\\ 1\\ 0\\ 0\\ 0\\ 0
\end{bmatrix}, e_{4-}^{(u)} = \begin{bmatrix}
0\\ 0\\ 1\\ 0\\ 0\\ 0\\ 0\\ 0\\ 0
\end{bmatrix},\\
 e_{5+}^{(u)} = \begin{bmatrix}
0\\ 0\\ 0\\ 0\\ 0\\ 0\\ 0\\ 0\\ 1
\end{bmatrix}, e_{6+}^{(u)} = \frac{1}{\sqrt{2}}\begin{bmatrix}
0\\ 0\\ 0\\ 0\\ 0\\ 0\\ 1\\ 1\\ 0
\end{bmatrix}, e_{7+}^{(u)} = \frac{1}{\sqrt{2}}\begin{bmatrix}
0\\ 1\\ 0\\ 0\\ 0\\ 1\\ 0\\ 0\\ 0
\end{bmatrix}, e_{8+}^{(u)} = \frac{1}{\sqrt{2}}\begin{bmatrix}
-1\\ 0\\ 0\\ 0\\ 1\\ 0\\ 0\\ 0\\ 0
\end{bmatrix}, e_{9+}^{(u)} = \begin{bmatrix}
0\\ 0\\ 0\\ 1\\ 0\\ 0\\ 0\\ 0\\ 0
\end{bmatrix},\\
e_{1-}^{(e)} = \frac{1}{\sqrt{2}}\begin{bmatrix}
0\\ 0\\ 0\\ 0\\ 0\\ 0\\ 0\\ -e^{i\frac{\sqrt{3}q_y}{2}}\\ 1
\end{bmatrix}, e_{2-}^{(e)} = \frac{1}{\sqrt{2}}\begin{bmatrix}
-1\\ 0\\ 0\\ 0\\ 0\\ 1\\ 0\\ 0\\ 0
\end{bmatrix}, e_{3-}^{(e)} = \frac{1}{\sqrt{2}}\begin{bmatrix}
0\\ -1\\ 0\\ 0\\ 1\\ 0\\ 0\\ 0\\ 0
\end{bmatrix}, e_{4-}^{(e)} = \frac{1}{\sqrt{2}}\begin{bmatrix}
0\\ 0\\ -1\\ 1\\ 0\\ 0\\ 0\\ 0\\ 0
\end{bmatrix},\\
e_{5+}^{(e)} = \frac{1}{\sqrt{2}}\begin{bmatrix}
0\\ 0\\ 0\\ 0\\ 0\\ 0\\ 0\\ e^{i\frac{\sqrt{3}q_y}{2}}\\ 1
\end{bmatrix}, e_{6+}^{(e)} = \begin{bmatrix}
0\\ 0\\ 0\\ 0\\ 0\\ 0\\ 1\\ 0\\ 0
\end{bmatrix}, e_{7+}^{(e)} = \frac{1}{\sqrt{2}}\begin{bmatrix}
1\\ 0\\ 0\\ 0\\ 0\\ 1\\ 0\\ 0\\ 0
\end{bmatrix}, e_{8+}^{(e)} = \frac{1}{\sqrt{2}}\begin{bmatrix}
0\\ 1\\ 0\\ 0\\ 1\\ 0\\ 0\\ 0\\ 0
\end{bmatrix}, e_{9+}^{(e)} = \frac{1}{\sqrt{2}}\begin{bmatrix}
0\\ 0\\ 1\\ 1\\ 0\\ 0\\ 0\\ 0\\ 0
\end{bmatrix},
\end{split}
\end{equation}
where the symbol $e^{(u/e)}_{i +/-}$ denotes $i$th eigenvector of $\mathbf{M}_{u/e}(q_x = 0,q_y)$ with eigenvalue $(+/-)1$. In the ordered basis $\{e^{(u)}_1,\dots,e^{(u)}_9\}$ and $\{e^{(e)}_1,\dots,e^{(e)}_9\}$, the matrix $\mathbf{C}(q_x=0,q_y)$ beceomes block-diagonal
\begin{equation}
\begin{split}
\tilde{\mathbf{C}}(q_x=0,q_y) &= \begin{pmatrix}
\mathbf{C}_{-}(q_x=0,q_y) & \mathbf{0}\\
 \mathbf{0} & \mathbf{C}_{+}(q_x=0,q_y)
\end{pmatrix},\\
\mathbf{C}_{-}(q_x=0,q_y) &= \begin{pmatrix}
0 & \frac{3}{2} & \frac{1}{2} & -\frac{1}{\sqrt{2}}e^{-i\frac{\sqrt{3}}{2}q_y}\\
\frac{6\delta}{\sqrt{3+36\delta^2}} & \frac{-3-6\delta}{2\sqrt{3+36\delta^2}} & \frac{-1+6\delta}{2\sqrt{1+12\delta^2}} & 0\\
\frac{-6\delta}{\sqrt{3+36\delta^2}} & 0 & 0 & \frac{-1-6\delta}{2\sqrt{2+24\delta^2}}\\
0 & \frac{-6\delta_3}{\sqrt{3+36\delta_3^2}} & \frac{1}{\sqrt{1+12\delta_3^2}} & 0
\end{pmatrix},\\
\mathbf{C}_{+}(q_x=0,q_y) &= \begin{pmatrix}
0 & 0 & \frac{3}{2} & \frac{1}{2} & -\sqrt{\frac{3}{2}}e^{-i\frac{\sqrt{3}}{2}q_y}\\
0 & 0 & 0 &\sqrt{2} & 0\\
0 & \frac{6\delta}{\sqrt{3+36\delta^2}} & \frac{-3-6\delta}{2\sqrt{3+36\delta^2}} & \frac{-1+6\delta}{2\sqrt{1+12\delta^2}} & 0\\
0 & \frac{6\delta}{\sqrt{3+36\delta^2}} & 0 & 0 & \frac{3-6\delta}{2\sqrt{6+72\delta^2}}\\
\frac{12\delta_3}{\sqrt{6+72\delta_3^2}} & 0 & \frac{6\delta_3}{\sqrt{3+36\delta_3^2}} & \frac{-1}{\sqrt{1+12\delta_3^2}} & 0
\end{pmatrix}.
\end{split}
\end{equation}
Determinant of these two matrices $\mathbf{C}_{-}(q_x=0,q_y)$ and $\mathbf{C}_{+}(q_x=0,q_y)$ for $\delta_3 = 1/3$ are:
\begin{subequations}
\label{Eq:detCpm}
\begin{align}
\det \left.\mathbf{C}_{+}^\dagger\right|_{L_1} &= \frac{6 \delta\sqrt{6/7}}{1+12\delta^2}\left((1-2\delta)+(1+2\delta)e^{i \sqrt{3}q_y/2}\right),\\
\det \left.\mathbf{C}_{-}^\dagger\right|_{L_1}&=- \frac{ \delta\sqrt{3/14}}{1+12\delta^2}\left(5(1+6\delta)+(5-6\delta)e^{i \sqrt{3}q_y/2}\right).
\end{align}
\end{subequations}

\section{$\mathbf{C}(\mathbf{q})$ can always be decomposed into $\mathbf{C}_+(\mathbf{q})$ and $\mathbf{C}_-(\mathbf{q})$ along the mirror invariant line}
To see this, we first note that along the mirror invariant line, due to mirror symmetry, we have
\begin{equation}
\label{eq:Csymmetry}
    \mathbf{M}_e^\dagger(\mathbf{q})\mathbf{C}(\mathbf{q})\mathbf{M}_u(\mathbf{q}) = \mathbf{C}(\mathbf{q}),
\end{equation}
by definition of mirror symmetry. Since reflecting twice about a mirror is identity, we have $\mathbf{M}_u^2(\mathbf{q})=\mathbbm{1}$ and $\mathbf{M}_e^2(\mathbf{q})=\mathbbm{1}$ along the mirror invariant line in Fourier space, and thus the eigenvalues of $\mathbf{M}_u(\mathbf{q})$ and $\mathbf{M}_e(\mathbf{q})$ are $\pm 1$. Let $\mathbf{M}_{u}(\mathbf{q})$ ($\mathbf{M}_{e}(\mathbf{q})$) have $n_{d+}$ ($n_{b+}$) eigenvectors $e_{1+}^{(u)},\dots,e_{n_{d+}+}^{(u)}$ ($e_{1+}^{(e)},\dots,e_{n_{b+}+}^{(e)}$) with eigenvalue $+1$, and $n_{d-}$ ($n_{b-}$) eigenvectors $e_{1-}^{(u)},\dots,e_{n_{d-}-}^{(u)}$ ($e_{1-}^{(e)},\dots,e_{n_{b-}-}^{(e)}$) with eigenvalue $-1$. Note that $n_{d+}+ n_{d-} = n_{d}$ is the total number of d.o.f.s in the unit cell, and $n_{b+}+ n_{b-} = n_{b}$ is the total number of bonds in the unit cell. Note that $e_{i+/-}^{(u)}$ ($e_{i+/-}^{(e)}$) are column vectors of size $n_d\times 1$ ($n_b\times 1$). These imply
\begin{subequations}
\begin{align}
    \mathbf{M}_u(\mathbf{q}) &= \left[e_{1+}^{(u)},\dots,e_{n_{d+}+}^{(u)},e_{1-}^{(u)},\dots,e_{n_{d-}-}^{(u)}\right] \begin{pmatrix}
        \mathbbm{1}_{n_{d+}\times n_{d+}} &\mathbf{0}_{n_{d+}\times n_{d-}}\\
        \mathbf{0}_{n_{d-}\times n_{d+}} & -\mathbbm{1}_{n_{d-}\times n_{d-}}
    \end{pmatrix}\begin{bmatrix}
            \left(e_{1+}^{(u)}\right)^\dagger\\
            \vdots\\
            \left(e_{{n_{d+}}+}^{(u)}\right)^\dagger\\
            \left(e_{1-}^{(u)}\right)^\dagger\\
            \vdots\\
            \left(e_{{n_{d-}}-}^{(u)}\right)^\dagger
        \end{bmatrix}\\
        &=\left[E_{+}^{(u)} E_{-}^{(u)}\right] \begin{pmatrix}
        \mathbbm{1}_{n_{d+}\times n_{d+}} &\mathbf{0}_{n_{d+}\times n_{d-}}\\
        \mathbf{0}_{n_{d-}\times n_{d+}} & -\mathbbm{1}_{n_{d-}\times n_{d-}}
    \end{pmatrix}\begin{bmatrix}
            {E_{+}^{(u)}}^\dagger\\
           {E_{-}^{(u)}}^\dagger
        \end{bmatrix},\\
        \mathbf{M}_e(\mathbf{q}) &= \left[e_{1+}^{(e)},\dots,e_{n_{b+}+}^{(e)},e_{1-}^{(e)},\dots,e_{n_{b-}-}^{(e)}\right] \begin{pmatrix}
        \mathbbm{1}_{n_{b+}\times n_{b+}} &\mathbf{0}_{n_{b+}\times n_{b-}}\\
        \mathbf{0}_{n_{b-}\times n_{b+}} & -\mathbbm{1}_{n_{b-}\times n_{b-}}
    \end{pmatrix}\begin{bmatrix}
            \left(e_{1+}^{(e)}\right)^\dagger\\
            \vdots\\
            \left(e_{{n_{b+}}+}^{(e)}\right)^\dagger\\
            \left(e_{1-}^{(e)}\right)^\dagger\\
            \vdots\\
            \left(e_{{n_{b-}}-}^{(e)}\right)^\dagger
        \end{bmatrix}
        \\
        &=\left[E_{+}^{(e)} E_{-}^{(e)}\right] \begin{pmatrix}
        \mathbbm{1}_{n_{b+}\times n_{b+}} &\mathbf{0}_{n_{b+}\times n_{b-}}\\
        \mathbf{0}_{n_{b-}\times n_{b+}} & -\mathbbm{1}_{n_{b-}\times n_{b-}}
    \end{pmatrix}\begin{bmatrix}
            {E_{+}^{(e)}}^\dagger\\
           {E_{-}^{(e)}}^\dagger
        \end{bmatrix},
\end{align}
\end{subequations}
where $E_{+/-}^{(u)} = \left[e_{1+/-}^{(u)},\dots,e_{n_{d+/-}+/-}^{(u)}\right]$ and $E_{+/-}^{(e)} = \left[e_{1+/-}^{(e)},\dots,e_{n_{b+/-}+/-}^{(u)}\right]$. Plugging these in Eq.~\eqref{eq:Csymmetry}, we obtain
\begin{subequations}
\begin{align}
    &\left[E_{+}^{(e)} E_{-}^{(e)}\right] \begin{pmatrix}
        \mathbbm{1}_{n_{b+}\times n_{b+}} &\mathbf{0}_{n_{b+}\times n_{b-}}\\
        \mathbf{0}_{n_{b-}\times n_{b+}} & -\mathbbm{1}_{n_{b-}\times n_{b-}}
    \end{pmatrix}\begin{bmatrix}
            {E_{+}^{(e)}}^\dagger\\
           {E_{-}^{(e)}}^\dagger
        \end{bmatrix}\mathbf{C}(\mathbf{q})\left[E_{+}^{(u)} E_{-}^{(u)}\right] \begin{pmatrix}
        \mathbbm{1}_{n_{d+}\times n_{d+}} &\mathbf{0}_{n_{d+}\times n_{d-}}\\
        \mathbf{0}_{n_{d-}\times n_{d+}} & -\mathbbm{1}_{n_{d-}\times n_{d-}}
    \end{pmatrix}\begin{bmatrix}
            {E_{+}^{(u)}}^\dagger\\
           {E_{-}^{(u)}}^\dagger
        \end{bmatrix} = \mathbf{C}(\mathbf{q})\nonumber\\
    &\Rightarrow \begin{pmatrix}
        \mathbbm{1}_{n_{b+}\times n_{b+}} &\mathbf{0}_{n_{b+}\times n_{b-}}\\
        \mathbf{0}_{n_{b-}\times n_{b+}} & -\mathbbm{1}_{n_{b-}\times n_{b-}}
    \end{pmatrix}\begin{bmatrix}
            {E_{+}^{(e)}}^\dagger\\
           {E_{-}^{(e)}}^\dagger
        \end{bmatrix}\mathbf{C}(\mathbf{q})\left[E_{+}^{(u)} E_{-}^{(u)}\right] \begin{pmatrix}
        \mathbbm{1}_{n_{d+}\times n_{d+}} &\mathbf{0}_{n_{d+}\times n_{d-}}\\
        \mathbf{0}_{n_{d-}\times n_{d+}} & -\mathbbm{1}_{n_{d-}\times n_{d-}}
    \end{pmatrix}\nonumber\\
    &\hspace{11cm}=\begin{bmatrix}
            {E_{+}^{(e)}}^\dagger\\
           {E_{-}^{(e)}}^\dagger
        \end{bmatrix}\mathbf{C}(\mathbf{q})\left[E_{+}^{(u)} E_{-}^{(u)}\right]\nonumber\\
    &\Rightarrow \begin{pmatrix}
        \mathbbm{1}_{n_{b+}\times n_{b+}} &\mathbf{0}_{n_{b+}\times n_{b-}}\\
        \mathbf{0}_{n_{b-}\times n_{b+}} & -\mathbbm{1}_{n_{b-}\times n_{b-}}
    \end{pmatrix} \begin{pmatrix}
        {E_{+}^{(e)}}^\dagger \mathbf{C}(\mathbf{q})E_{+}^{(u)} &{E_{+}^{(e)}}^\dagger \mathbf{C}(\mathbf{q})E_{-}^{(u)}\\
        {E_{-}^{(e)}}^\dagger \mathbf{C}(\mathbf{q})E_{+}^{(u)} & {E_{-}^{(e)}}^\dagger \mathbf{C}(\mathbf{q})E_{-}^{(u)}
    \end{pmatrix}\begin{pmatrix}
        \mathbbm{1}_{n_{d+}\times n_{d+}} &\mathbf{0}_{n_{d+}\times n_{d-}}\\
        \mathbf{0}_{n_{d-}\times n_{d+}} & -\mathbbm{1}_{n_{d-}\times n_{d-}}
    \end{pmatrix}\nonumber\\
    &\hspace{11cm}=\begin{pmatrix}
        {E_{+}^{(e)}}^\dagger \mathbf{C}(\mathbf{q})E_{+}^{(u)} &{E_{+}^{(e)}}^\dagger \mathbf{C}(\mathbf{q})E_{-}^{(u)}\\
        {E_{-}^{(e)}}^\dagger \mathbf{C}(\mathbf{q})E_{+}^{(u)} & {E_{-}^{(e)}}^\dagger \mathbf{C}(\mathbf{q})E_{-}^{(u)}
    \end{pmatrix}\nonumber\\
    &\Rightarrow \begin{pmatrix}
        {E_{+}^{(e)}}^\dagger \mathbf{C}(\mathbf{q})E_{+}^{(u)} &-{E_{+}^{(e)}}^\dagger \mathbf{C}(\mathbf{q})E_{-}^{(u)}\\
        -{E_{-}^{(e)}}^\dagger \mathbf{C}(\mathbf{q})E_{+}^{(u)} & {E_{-}^{(e)}}^\dagger \mathbf{C}(\mathbf{q})E_{-}^{(u)}
    \end{pmatrix}=\begin{pmatrix}
        {E_{+}^{(e)}}^\dagger \mathbf{C}(\mathbf{q})E_{+}^{(u)} &{E_{+}^{(e)}}^\dagger \mathbf{C}(\mathbf{q})E_{-}^{(u)}\\
        {E_{-}^{(e)}}^\dagger \mathbf{C}(\mathbf{q})E_{+}^{(u)} & {E_{-}^{(e)}}^\dagger \mathbf{C}(\mathbf{q})E_{-}^{(u)}
    \end{pmatrix}\nonumber\\
    &\Rightarrow {E_{+}^{(e)}}^\dagger \mathbf{C}(\mathbf{q})E_{-}^{(u)} = \mathbf{0},\,\, {E_{-}^{(e)}}^\dagger \mathbf{C}(\mathbf{q})E_{+}^{(u)} = \mathbf{0},
\end{align}
\end{subequations}
and consequently, in the eigenbasis of $\mathbf{M}_u(\mathbf{q})$ and $\mathbf{M}_e(\mathbf{q})$, the compatibility matrix has form
\begin{equation}
    \tilde{C}(\mathbf{q}) = \begin{bmatrix}
            {E_{+}^{(e)}}^\dagger\\
           {E_{-}^{(e)}}^\dagger
        \end{bmatrix}\mathbf{C}(\mathbf{q})\left[E_{+}^{(u)} E_{-}^{(u)}\right] = \begin{pmatrix}
        {E_{+}^{(e)}}^\dagger \mathbf{C}(\mathbf{q})E_{+}^{(u)} &\mathbf{0}\\
        \mathbf{0} & {E_{-}^{(e)}}^\dagger \mathbf{C}(\mathbf{q})E_{-}^{(u)}
    \end{pmatrix},
\end{equation}
and we identify $\mathbf{C}_{+/-}(\mathbf{q}) = {E_{+/-}^{(e)}}^\dagger \mathbf{C}(\mathbf{q})E_{+/-}^{(u)}$; these two matrices have sizes $n_{b+/-}\times n_{d+/-}$. This is result is very general and always true as long as there is a mirror symmetry. When $n_{b+} = n_{d-}$ (and consequently $n_{b-} = n_{d-}$, since it is Maxwell frame), these two matrices are square matrices and one can evaluate the determinants of them. 

\section{Regime where $\mathcal{H}(\mathbf{q})$ or $\mathbf{D}(\mathbf{q})$ is fully gapped at $\omega = 0$}\label{Sec:FullyGapped}
\begin{figure}[h]
\includegraphics{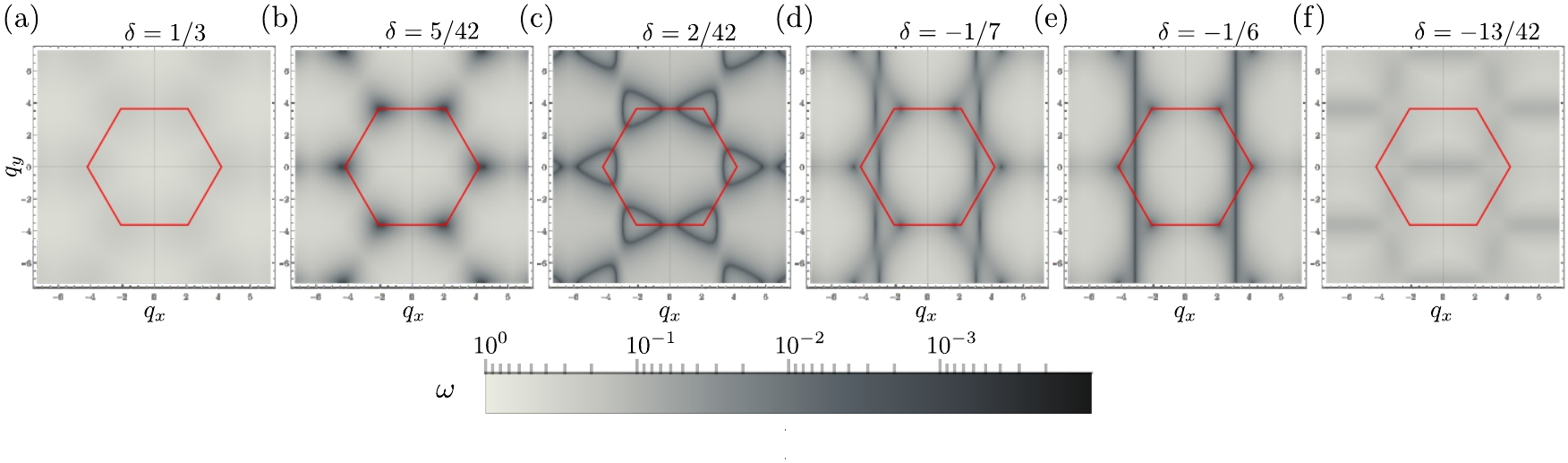}
\caption{Density plot of the spectrum of lowest frequency band of $\mathbf{D}(\mathbf{q})$ for different values of $\delta$. The red hexagon shows the edge of the Brillouin zone. The corners of the red hexagon are the $K/K'$ points.  Bands in (b-e) touch $\omega = 0$ whereas bands in (a) and (f) are never touch $\omega = 0$.}
\label{Fig:LowestBand}
\end{figure}
For simplicity, let us keep $\delta_3 = 1/3$ and vary $\delta_1 = \delta_2 \equiv \delta$. When $\delta = 0$, there is no restoring force to the linear order to the displacements of points $4$ and $5$. Therefore, at $\delta = 0$, there are two completely flat zero frequency bands of the dynamical matrix $\mathbf{D}(\mathbf{q})$ and four flat zero frequency bands of the dynamical matrix $\mathcal{H}(\mathbf{q})$. As we increase $\delta$ from $\delta = 0$, we see rings of zero frequency Weyl lines surrounding $K$ and $K'$ point in the Brillouin zone ($\Gamma$, $K$, $K'$ and $M'$ are gapped) as can be seen in Fig.~\ref{Fig:LowestBand}(c). However, as we increase $\delta$, these rings get tighter around the $K$ and $K'$ point, and at some value of $\delta$ they disappear giving a band structure completely gapped at $\omega = 0$. Then, clearly at the transition point between gapless and gapped zero frequency, there are zero modes at $K$ and $K'$ points. In other words, we have to find the value of $\delta$ for which the $\det(\mathbf{C}(q_x = 4\pi/3,q_y = 0)) = 0$. At $K$ point,
\begin{equation}
\det(\mathbf{C}(q_x = 4\pi/3,q_y = 0)) = \frac{9\delta^2}{2(1+12\delta^2)^2}(1+6\delta)(5-42\delta).
\end{equation}
There is a zero mode at $K'$ point when $\delta = 5/42$ (see Fig.~\ref{Fig:LowestBand}(b)). This implies that the band structure is fully gapped at $\delta>5/42$ (Fig.~\ref{Fig:LowestBand}(a)). Similarly, for $\delta<0$, there are isolated zero modes on the line $M'-K'$ as well as vertical lines of zero modes on either side of $q_x = \pi$ (see Fig.~\ref{Fig:LowestBand}(d)). They disappear after the isolated zero modes hit point $K$ and the vertical lines of zero mode hit $q_x = \pi$. From the above equation, we see that the first one happens at $\delta=-1/6$ (see Fig.~\ref{Fig:LowestBand}(e)). For the latter, we calculate $\det(\mathbf{C}(q_x = \pi,q_y = 0))$:
\begin{equation}
\det(\mathbf{C}(q_x = 4\pi/3,q_y = 0)) = -\frac{288\delta^3}{7(1+12\delta^2)^2}(1+6\delta).
\end{equation}
Hence, there is a line of zero mode at $q_x = \pi$ for $\delta = -1/6$ (see Fig.~\ref{Fig:LowestBand}(e)). When $\delta<-1/6$, $\omega = 0$ is gapped everywhere in the Brillouin zone (see Fig.~\ref{Fig:LowestBand}(f)). However, the band gap closes again at the $M'$ point at $\delta = -5/12$ which can be seen from Eq.~(\ref{Eq:detCpm}b).

\section{Low energy theory and the edge states at mirror invariant domain walls}\label{Sec:LowEnergyDW}
Since, $\delta = 0$ is the phase transition point between phase 1 and phase 2, we can write a low energy theory near $\delta = 0$ near the $M'$ point: $\mathbf{q} = (0,2\pi/\sqrt{3}) + (k_x,k_y)$ keeping only lowest few orders in the momenta $k_x$ and $k_y$. Note that we are expanding near the $M'$ point because along the line of our interest $\Gamma-M'$, the gap is smallest near the $M'$ point for small values of $\delta$ (see Figs.~\sss{2}(b) and (d)).
Using this low energy theory, we can explicitly show the existence of edge modes at a domain wall between phase $\delta>0$ and $\delta<0$. Using the low energy method to show existence of boundary mode is known as Jackiw-Rebbi analysis in the literature~\citesupp{JackiwRebbi1,BradlynCornerRobustness1}.

\subsection{Integrate the high frequency bands to obtain the low energy theory}
To get to the low energy theory with small $\delta$, $k_x$  and $k_y$, we first do a singular value decomposition at the matrix $\mathbf{C}(\mathbf{q} = (0,2\pi/\sqrt{3}))|_{\delta = 0}$:
\begin{equation}
\mathbf{C}(\mathbf{q} = (0,2\pi/\sqrt{3}))|_{\delta = 0} = \mathbf{U} \begin{pmatrix}
\mathbf{\Lambda}_{7\times 7} & \mathbf{0}_{7\times 2}\\
\mathbf{0}_{2\times 7} & \mathbf{0}_{2\times 2}
\end{pmatrix} \mathbf{W}^\dagger, 
\end{equation}
where $\mathbf{\Lambda}$ is diagonal matrix consisting of the 7 nonzero singular values. These are the nonzero finite frequencies at the $M'$ point for $\delta = 0$. There are two 0 singular values since there are no restoring force to points 4 and 5. The columns of the matrix $\mathbf{W}$ are the eigenvectors of $\mathbf{D}(\mathbf{q} = (0,2\pi/\sqrt{3}))|_{\delta = 0}$ whereas the columns of $\mathbf{U}$ are the eigenvectors of $\left[\mathbf{C}(\mathbf{q} = (0,2\pi/\sqrt{3}))\mathbf{C}^\dagger(\mathbf{q} = (0,2\pi/\sqrt{3}))\right]_{\delta=0}$. The matrices $\mathbf{U}$ and $\mathbf{W}$ are of the form:
\begin{equation}
\mathbf{U} = \left[\mathbf{U}_H | \mathbf{U}_L\right], \mathbf{W} = \left[\mathbf{W}_H | \mathbf{W}_L\right]
\end{equation}
where $\mathbf{U}_H$ ($\mathbf{W}_H$) is $9\times 7$ matrix containing the 7 eigenvectors of $\left[\mathbf{C}(\mathbf{q} = (0,2\pi/\sqrt{3}))\mathbf{C}^\dagger(\mathbf{q} = (0,2\pi/\sqrt{3}))\right]_{\delta=0}$ ($\mathbf{D}(\mathbf{q} = (0,2\pi/\sqrt{3}))|_{\delta = 0}$) corresponding to the nonzero eigenvalues. The matrices $\mathbf{U}_L$ and $\mathbf{W}_L$ are $9\times 2$ containing 2 eigenvectors corresponding to 0 eigenvalues.

Now, we ask how the elements of this matrix are changed once we allow small $\delta$, $k_x$ and $k_y$. To facilitate this expansion, we multiply a small parameter $\varepsilon$ to $\delta$, $k_x$ and $k_y$ and expand the matrix in Taylor series of $\varepsilon$:
\begin{equation}
\begin{split}
\mathbf{C}(\mathbf{q} = (0,2\pi/\sqrt{3}) + \varepsilon(k_x,k_y))|_{\delta\rightarrow \varepsilon \delta} &= \mathbf{C}(\mathbf{q} = (0,2\pi/\sqrt{3}))|_{\delta = 0} +\varepsilon \mathbf{C}_1 + \varepsilon^2 \mathbf{C}_2 +\mathcal{O}(\varepsilon^3)\\
& = \begin{pmatrix}\mathbf{U}_H & \mathbf{U}_L\end{pmatrix}
\begin{pmatrix}
\mathbf{\Lambda}+\varepsilon \mathbf{P}_1+\varepsilon^2 \mathbf{P}_2 +\mathcal{O}(\varepsilon^3) & \varepsilon \mathbf{Q}_1+\varepsilon^2 \mathbf{Q}_2 +\mathcal{O}(\varepsilon^3)\\
\varepsilon \mathbf{R}_1+\varepsilon^2 \mathbf{R}_2 +\mathcal{O}(\varepsilon^3) & \varepsilon \mathbf{S}_1+\varepsilon^2 \mathbf{S}_2 +\mathcal{O}(\varepsilon^3)
\end{pmatrix} \begin{pmatrix}
\mathbf{W}_H^\dagger\\
\mathbf{W}_L^\dagger
\end{pmatrix},
\end{split}
\end{equation}
where
\begin{equation}
\mathbf{P}_i = \mathbf{U}_H^\dagger \mathbf{C}_i  \mathbf{W}_H, \mathbf{Q}_i = \mathbf{U}_H^\dagger \mathbf{C}_i  \mathbf{W}_L, \mathbf{R}_i = \mathbf{U}_L^\dagger \mathbf{C}_i  \mathbf{W}_H, \mathbf{S}_i = \mathbf{U}_L^\dagger \mathbf{C}_i  \mathbf{W}_L.
\end{equation}
Now, our aim is to integrate out the finite frequency modes and keep the low energy mode. Since the columns of $\mathbf{W}$ form a complete basis for the displacements, we can write any displacement in this basis as $\mathbf{u} = (\mathbf{u}_H^\dagger, \mathbf{u}_L^\dagger)^\dagger$, where $\mathbf{u}_H$ ($\mathbf{u}_L$) contain the amplitudes of the high (low) frequency modes. The energy of the system is:
\begin{equation}
\begin{split}
E &= \begin{pmatrix}\mathbf{u}_H^\dagger & \mathbf{u}_L^\dagger\end{pmatrix} \begin{pmatrix}
\mathbf{A}^\dagger & \mathbf{C}^\dagger\\
\mathbf{B}^\dagger & \mathbf{E}^\dagger
\end{pmatrix} \begin{pmatrix}
\mathbf{A} & \mathbf{B}\\
\mathbf{C} & \mathbf{E}
\end{pmatrix} \begin{pmatrix}\mathbf{u}_H\\ \mathbf{u}_L\end{pmatrix}\\
&=\mathbf{u}_H^\dagger (\mathbf{A}^\dagger \mathbf{A}+\mathbf{C}^\dagger \mathbf{C}) \mathbf{u}_H + \mathbf{u}_H^\dagger (\mathbf{A}^\dagger \mathbf{B}+\mathbf{C}^\dagger \mathbf{E}) \mathbf{u}_L+\mathbf{u}_L^\dagger (\mathbf{B}^\dagger \mathbf{A}+\mathbf{E}^\dagger \mathbf{C}) \mathbf{u}_L + \mathbf{u}_L^\dagger (\mathbf{B}^\dagger \mathbf{B}+\mathbf{E}^\dagger \mathbf{E}) \mathbf{u}_L\\
&= (\mathbf{u}_H + (\mathbf{A}^\dagger \mathbf{A}+\mathbf{C}^\dagger \mathbf{C})^{-1}(\mathbf{A}^\dagger \mathbf{B}+\mathbf{C}^\dagger \mathbf{E}) \mathbf{u}_L)^\dagger  (\mathbf{A}^\dagger \mathbf{A}+\mathbf{C}^\dagger \mathbf{C}) (\mathbf{u}_H + (\mathbf{A}^\dagger \mathbf{A}+\mathbf{C}^\dagger \mathbf{C})^{-1}(\mathbf{A}^\dagger \mathbf{B}+\mathbf{C}^\dagger \mathbf{E}) \mathbf{u}_L)\\
&\phantom{= }+ \mathbf{u}_L^\dagger (\mathbf{B}^\dagger \mathbf{B}+\mathbf{E}^\dagger \mathbf{E} - (\mathbf{A}^\dagger \mathbf{B}+\mathbf{C}^\dagger \mathbf{E})^\dagger (\mathbf{A}^\dagger \mathbf{A}+\mathbf{C}^\dagger \mathbf{C})^{-1} (\mathbf{A}^\dagger \mathbf{B}+\mathbf{C}^\dagger \mathbf{E})) \mathbf{u}_L,
\end{split}
\end{equation}
where $\mathbf{A} = \mathbf{\Lambda}+\varepsilon \mathbf{P}_1+\varepsilon^2 \mathbf{P}_2 +\mathcal{O}(\varepsilon^3)$, $\mathbf{B} = \varepsilon \mathbf{Q}_1+\varepsilon^2 \mathbf{Q}_2 +\mathcal{O}(\varepsilon^3)$, $\mathbf{C} = \varepsilon \mathbf{R}_1+\varepsilon^2 \mathbf{R}_2 +\mathcal{O}(\varepsilon^3)$ and $\mathbf{E} = \varepsilon \mathbf{S}_1+\varepsilon^2 \mathbf{S}_2 +\mathcal{O}(\varepsilon^3)$. Since $\mathbf{A}^\dagger \mathbf{A}+\mathbf{C}^\dagger \mathbf{C} = \mathbf{\Lambda}^2+\mathcal{O}(\varepsilon)$ and $\mathbf{\Lambda}$ is a diagonal matrix with nonzero finite entries in the diagonal, $\mathbf{A}^\dagger \mathbf{A}+\mathbf{C}^\dagger \mathbf{C} $ is invertible. The effective low energy dynamical matrix is then
\begin{equation}
\mathbf{D}_L= \mathbf{B}^\dagger \mathbf{B}+\mathbf{E}^\dagger \mathbf{E} - (\mathbf{A}^\dagger \mathbf{B}+\mathbf{C}^\dagger \mathbf{E})^\dagger (\mathbf{A}^\dagger \mathbf{A}+\mathbf{C}^\dagger \mathbf{C})^{-1} (\mathbf{A}^\dagger \mathbf{B}+\mathbf{C}^\dagger \mathbf{E}).
\end{equation}
We will expand this expression in orders of $\varepsilon$.
\begin{equation}
\begin{split}
\mathbf{B}^\dagger \mathbf{B}+\mathbf{E}^\dagger \mathbf{E} &= \varepsilon^2(\mathbf{Q}_1^\dagger \mathbf{Q}_1+\mathbf{S}_1^\dagger \mathbf{S}_1) + \varepsilon^3(\mathbf{Q}_1^\dagger \mathbf{Q}_2 + \mathbf{Q}_2^\dagger \mathbf{Q}_1+\mathbf{S}_1^\dagger \mathbf{S}_2+\mathbf{S}_2^\dagger \mathbf{S}_1) +\mathcal{O}(\varepsilon^4),\\
\mathbf{A}^\dagger \mathbf{B}+\mathbf{C}^\dagger \mathbf{E} &= \varepsilon (\mathbf{\Lambda}^\dagger\mathbf{Q}_1)+\varepsilon^2(\mathbf{P}_1^\dagger \mathbf{Q}_1+\mathbf{\Lambda}^\dagger \mathbf{Q}_2+\mathbf{R}_1^\dagger \mathbf{S}_1) +\mathcal{O}(\varepsilon^3),\\
(\mathbf{A}^\dagger \mathbf{A}+\mathbf{C}^\dagger \mathbf{C})^{-1} &= (\mathbf{\Lambda}^\dagger\mathbf{\Lambda}+\varepsilon (\mathbf{\Lambda}^\dagger\mathbf{P}_1+\mathbf{P}_1^\dagger\mathbf{\Lambda})+\mathcal{O}(\varepsilon^2))^{-1}\\
&= (\mathbf{\Lambda}^\dagger\mathbf{\Lambda})^{-1} - \varepsilon(\mathbf{P}_1\mathbf{\Lambda}^{-1}+(\mathbf{\Lambda}^\dagger)^{-1}\mathbf{P}_1^\dagger)+\mathcal{O}(\varepsilon^2).
\end{split}
\end{equation}
Using these we get:
\begin{equation}
\begin{split}
\mathbf{D}_L &= \mathbf{B}^\dagger \mathbf{B}+\mathbf{E}^\dagger \mathbf{E} - (\mathbf{A}^\dagger \mathbf{B}+\mathbf{C}^\dagger \mathbf{E})^\dagger (\mathbf{A}^\dagger \mathbf{A}+\mathbf{C}^\dagger \mathbf{C})^{-1} (\mathbf{A}^\dagger \mathbf{B}+\mathbf{C}^\dagger \mathbf{E})\\
&= \varepsilon^2(\mathbf{Q}_1^\dagger \mathbf{Q}_1+\mathbf{S}_1^\dagger \mathbf{S}_1) + \varepsilon^3(\mathbf{Q}_1^\dagger \mathbf{Q}_2 + \mathbf{Q}_2^\dagger \mathbf{Q}_1+\mathbf{S}_1^\dagger \mathbf{S}_2+\mathbf{S}_2^\dagger \mathbf{S}_1)\\
 &\phantom{=} - \varepsilon^2\mathbf{Q}_1^\dagger \mathbf{Q}_1 -\varepsilon^3(\mathbf{Q}_1^\dagger \mathbf{\Lambda}^{-1} \mathbf{P}_1^\dagger \mathbf{Q}_1+\mathbf{Q}_1^\dagger \mathbf{Q}_2+\mathbf{Q}_1^\dagger \mathbf{\Lambda}^{-1} \mathbf{R}_1^\dagger \mathbf{S}_1)\\
&\phantom{=} -\varepsilon^3(\mathbf{Q}_1^\dagger \mathbf{P}_1\mathbf{\Lambda}^{-1}\mathbf{Q}_1+\mathbf{Q}_2^\dagger \mathbf{Q}_1+\mathbf{S}_1^\dagger \mathbf{R}_1\mathbf{\Lambda}^{-1} \mathbf{Q}_1)\\
&\phantom{=} +\varepsilon^3(\mathbf{Q}_1^\dagger \mathbf{P}_1\mathbf{\Lambda}^{-1}\mathbf{Q}_1+\mathbf{Q}_1^\dagger \mathbf{\Lambda}^{-1} \mathbf{P}_1^\dagger \mathbf{Q}_1)+\mathcal{O}(\varepsilon^4)\\
&= \varepsilon^2\mathbf{S}_1^\dagger \mathbf{S}_1 + \varepsilon^3(\mathbf{S}_1^\dagger(\mathbf{S}_2- \mathbf{R}_1\mathbf{\Lambda}^{-1} \mathbf{Q}_1)+(\mathbf{S}_2- \mathbf{R}_1\mathbf{\Lambda}^{-1} \mathbf{Q}_1)^\dagger\mathbf{S}_1)+\mathcal{O}(\varepsilon^4)\\
&= (\varepsilon\mathbf{S}_1+\varepsilon^2(\mathbf{S}_2- \mathbf{R}_1\mathbf{\Lambda}^{-1} \mathbf{Q}_1))^\dagger(\varepsilon\mathbf{S}_1+\varepsilon^2(\mathbf{S}_2- \mathbf{R}_1\mathbf{\Lambda}^{-1} \mathbf{Q}_1))+\mathcal{O}(\varepsilon^4).
\end{split}
\end{equation}
Therefore, the effective compatibility matrix in the low-energy sector is
\begin{equation}
\mathbf{C}_L = \varepsilon\mathbf{S}_1+\varepsilon^2(\mathbf{S}_2- \mathbf{R}_1\mathbf{\Lambda}^{-1} \mathbf{Q}_1)+\mathcal{O}(\varepsilon^3).
\end{equation}
Using this formula and definitions of $\mathbf{Q}_i$, $\mathbf{R}_i$, $\mathbf{S}_i$ and $\mathbf{\Lambda}$ from above, the effective compatibility matrix in the low energy sector for our system is evaluated to
\begin{equation}
\mathbf{C}_L(\mathbf{q} = (0,2\pi/\sqrt{3}) + (k_x,k_y))|_\delta = \begin{pmatrix}
11.2 \delta^2 & 3.2\delta^2 +i\delta(k_x+\sqrt{3}k_y)\\
-3.2\delta^2 +i\delta(k_x-\sqrt{3}k_y) & -11.2 \delta^2
\end{pmatrix}.
\end{equation}
Note that this matrix $\mathbf{C}_L$ is written in the basis $\mathbf{U}_L = \{1/\sqrt{3}\{0,-1,0,0,0,1,0,0,1\}^T,1/\sqrt{3}\{1,0,0,0,-1,0,0,-1,0\}^T\}$ and $\mathbf{W}_L =  \{\{0,0,0,0,0,0,-1,0,0\}^T,\{0,0,0,0,0,0,0,1,0\}^T\}$. In these bases, the mirror operators are
\begin{equation}
\begin{split}
\mathbf{M}_u^{L} = \begin{pmatrix}
0 & -1\\
-1 & 0
\end{pmatrix}, &\;\mathbf{M}_e^{L} = \begin{pmatrix}
0& 1\\
1 & 0
\end{pmatrix}, \\
\mathbf{M}_e^{L}\mathbf{C}_L(\mathbf{q} = (0,2\pi/\sqrt{3}) + (k_x,k_y))|_\delta \mathbf{M}_u^{L} &= \mathbf{C}_L(\mathbf{q} = (0,2\pi/\sqrt{3}) + (-k_x,k_y))|_\delta
\end{split}
\end{equation}
Moreover, when $\delta$ is zero, the whole matrix is zero meaning the eigenvalues are zero for all $k_x$ and $k_y$. This agrees with the full dynamical matrix where we saw that the lowest bands are zero when $\delta = 0$.
\subsection{Zero frequency edge modes at domain wall from the low energy theory}
Now, to create a domain wall between phase 1 and 2 at $y=0$ we have two choices:
\begin{enumerate}
\item $\delta<0$ when $y<0$, $\delta>0$ when $y>0$,
\item $\delta>0$ when $y<0$, $\delta<0$ when $y>0$.
\end{enumerate}
We will consider these two cases separately. Note that now we have to replace $k_y$ with $-i\partial_y$ since the translation  symmetry is broken in the $y$-direction.

\begin{figure}[t]
\includegraphics{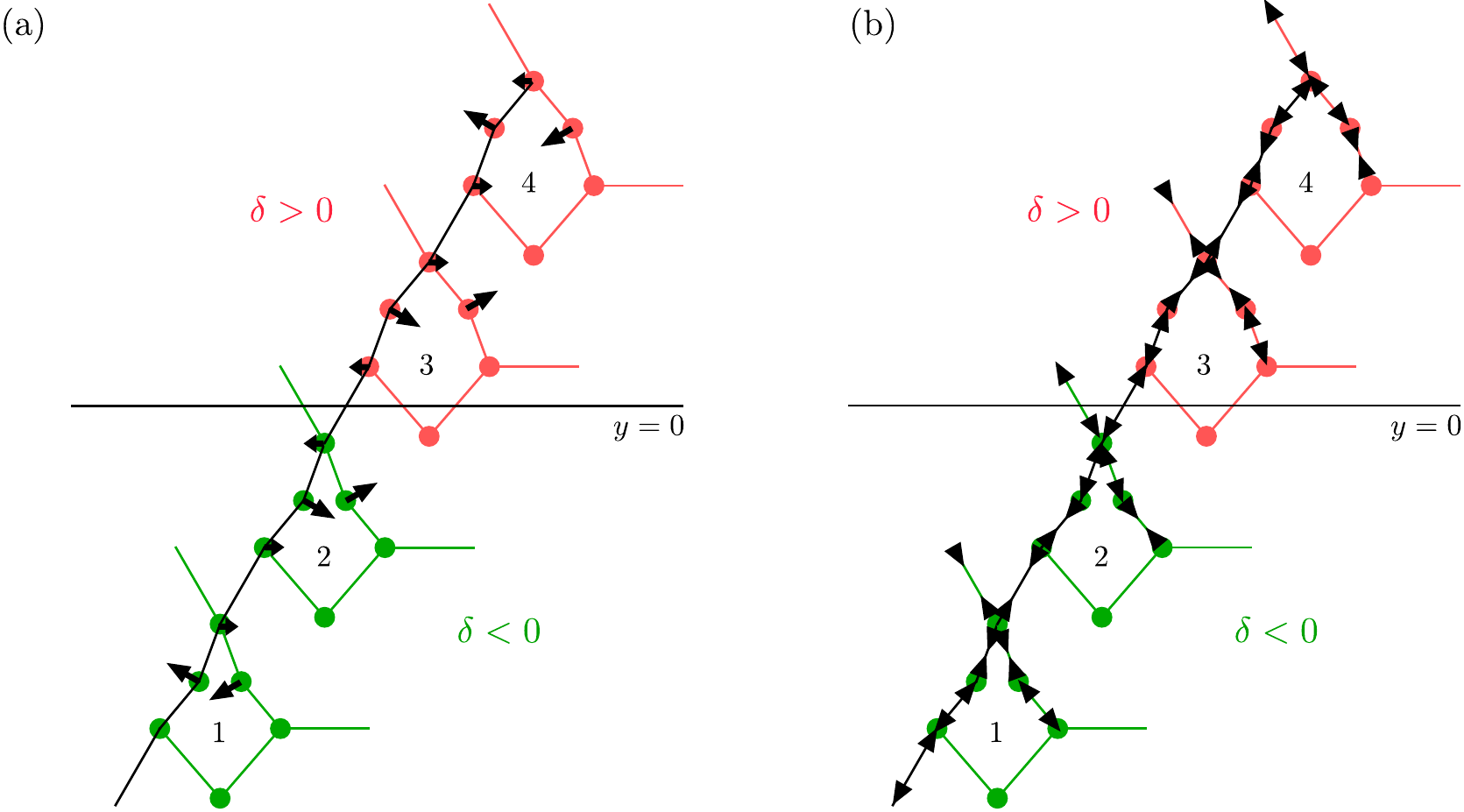}
\caption{Demonstration of boundary conditions for Eqs.~\eqref{Eq:fbu} and~\eqref{Eq:fbe}. In (a-b), a domain wall between the two phases $\delta>0$ and $\delta<0$ is considered. The unit cells are enumerated in increasing order from bottom to top. In (a) starting from opposite displacements of 4th and 5th node of unit cell 1, the directions of displacements of all other nodes are shown  such that the bonds are not elongation. In (b) starting from tensions in the bonds of unit cell 1, the tensions in all other bonds are shown such that there is no force on nodes 1-3 and only nonzero force perpendicular to displacement directions of nodes 4 and 5.}
\label{Fig:figureSb}
\end{figure}

\underline{Case 1: $\text{sgn}(\delta) = \text{sgn}(y)$} This is the case at the bottom domain wall in Fig.~\sss{3} of the main text. We will start by showing that there is a zero mode of the compatibility matrix $\mathbf{C}_L$ at $k_x = 0$. We choose the form of the zero mode to be $\psi_b^u(y) =f_b^u(y)(a,b)^T$, where $a$ and $b$ scalar numbers and the $y$-dependence is captured in the function $f_b^u(y)$. In other words, we seek a solution to the following problem
\begin{equation}
\begin{split}
\mathbf{C}_L(k_x = 0,k_y \rightarrow -i\partial_y)\psi_b^u(y) &=\mathbf{0}\\
\Rightarrow
\begin{pmatrix}
11.2 \delta^2 & 3.2\delta^2 +\delta\sqrt{3}\partial_y\\
-3.2\delta^2 -\delta\sqrt{3}\partial_y & -11.2 \delta^2
\end{pmatrix}f_b^u(y)\begin{pmatrix}a\\ b\end{pmatrix}&= \begin{pmatrix}0\\ 0\end{pmatrix}\\
\Rightarrow [11.2  \delta^2 \sigma_z + i( 3.2 \delta^2+\sqrt{3}\delta \partial_y) \sigma_y] f_b^u(y)\begin{pmatrix}a\\ b\end{pmatrix}&=\mathbf{0}.
\end{split}
\end{equation}
Note that we want the function $f_b^u(y)$ to be localized at $y=0$, i.e., exponentially decaying away from $y = 0$. Multiplying by $\sigma_x$ from the left on both sides, we get
\begin{equation}
[11.2  \delta^2\mathbbm{1} + ( 3.2 \delta^2+\sqrt{3}\delta \partial_y) \sigma_x] f_b^u(y)\begin{pmatrix}a\\ b\end{pmatrix}=\mathbf{0}.
\end{equation}
Choosing $(a,b) = (1,1)/\sqrt{2}$, the equation becomes a scalar first order differential equation
\begin{equation}
\label{Eq:fbu}
\sqrt{3} \partial_y f_b^u(y) + 14.4 \delta f_b^u(y) = 0 \Rightarrow f_b^u(y) = \begin{cases}c_{1>}^u e^{-14.4/\sqrt{3} \int_0^y dy'\delta(y')}, & y>0\\
c_{1<}^u e^{-14.4/\sqrt{3} \int_0^y dy'\delta(y')},  & y<0,
\end{cases}
\end{equation}
where $c_{1>}^u$ and $c_{1<}^u$ are constants of integration. To find the relation between these two constants, we have to use appropriate boundary condition. The claim is that $c_{1>}^u = -c_{1<}^u$. To see this we first recall that we wrote the low energy theory around $\mathbf{q} = (0,2\pi/\sqrt{3})$, as a result the sign of the displacements changes from one unit cell to the next in the direction $(1/2,\sqrt{3}/2)$. Moreover, the low energy theory was written in the displacement basis $\mathbf{W}_L =  \{\{0,0,0,0,0,0,-1,0,0\}^T,\{0,0,0,0,0,0,0,1,0\}^T\}$, and we found that on each side of the domain wall the zero mode is $(1,1)/\sqrt{2}$ in this basis. Therefore, in each unit cell the $7$th and $8$th degrees of freedom (displacements of 4th and 5th node in the unit cell as shown in Fig.~\sss{1}(a) of main text) have displacement of opposite sign. With this information, we turn to Fig.~\ref{Fig:figureSb}(a). In unit cell 1 of Fig.~\ref{Fig:figureSb}(a), the displacements of 4th and 5th nodes are shown. If bonds 2 and 5 (see Fig.~\sss{1}(a) of main text) are to be in their equilibrium length, node 2 need to be displaced by a small amount in the shown direction. Then node 3 of unit cell 2 need to move in the same direction by the same amount for bond 9 of unit cell 2 to be at its equilibrium length. Now, for bond 6 of unit cell to be of equilibrium length, node 5 of unit cell 2 clearly need to be displaced in the opposite direction to that of unit cell 1. This confirms the change of sign from one unit cell to the next in direction $(1/2,\sqrt{3}/2)$ as was predicted before from the low energy theory around $\mathbf{q} = (0,2\pi/\sqrt{3})$. However, following this procedure up to the 3rd unit cell, we see that the displacement of the 4th and 5th nodes of the 3rd unit cell are in the same direction as those of the 2nd unit cell. To get this same sign between 2nd unit cell ($y<0$) to the 3rd unit cell ($y>0$) on top of the effect of $\mathbf{q} = (0,2\pi/\sqrt{3})$, we need $c_{1>}^u = -c_{1<}^u$. In a compact form, we can then write
\begin{equation}
    f_b^u(y) = c_{1}^u \text{sgn}(y) e^{-14.4/\sqrt{3}\int_0^y dy'\delta(y')}.
\end{equation}
Here $c_1^u$ is a constant chosen to normalize zero mode $\psi_b^u(y)$. The function $f_b^u(y)$ is exponentially decaying away from $y = 0$ due to the fact that $\delta>0$ for $y>0$ and $\delta<0$ for $y<0$. Therefore, zero frequency edge mode of $\mathbf{C}_L(k_x = 0)$ is $\psi_b^u(y) = c_1^u \text{sgn}(y) e^{-14.4/\sqrt{3} \int_0^y dy'\delta(y')}(1,1)^T/\sqrt{2}$. Note that we could have chosen $(a,b) = (1,-1)/\sqrt{2}$, but in that case the differential equation would be $\sqrt{3} \partial_y f_b^u(y) - 8 \delta f_b^u(y) = 0$ which does not have an exponentially localized solution near $y = 0$. We can check how the zero frequency edge mode $\psi_b^u(y)$ transforms under the effective mirror operator $\mathbf{M}_u^L$:
\begin{equation}
\mathbf{M}_u^L \psi_b^u(y) = \begin{pmatrix}0& -1\\-1 & 0 \end{pmatrix}\frac{f_b^u(y)}{\sqrt{2}}\begin{pmatrix}1\\1 \end{pmatrix} = -\frac{f_b^u(y)}{\sqrt{2}}\begin{pmatrix}1\\1 \end{pmatrix} = -\psi_b^u(y),
\end{equation}
meaning $\psi_b(y)$ is odd under mirror $m_x$. This matches with the plot in Fig.~\sss{3}(c).

Next we find the state of self stress at this domain wall. For that we will work with the matrix $\mathbf{C}_L^\dagger$:
\begin{equation}
\mathbf{C}_L^\dagger = 11.2  \delta^2 \sigma_z - i\delta k_x \sigma_x + ( -3.2 i \delta^2-\sqrt{3}\delta k_y) \sigma_y \rightarrow 11.2  \delta^2 \sigma_z - i\delta k_x \sigma_x + ( -3.2 i \delta^2+i\sqrt{3}\delta\partial_y) \sigma_y
\end{equation}
Similar to before, we are going to consider solution of the form $\psi_b^e(y) =f_b^e(y)(a,b)^T$ with exponentially localized $f_b^e(y)$. Setting $k_x = 0$ and multiplying by $\sigma_z$ from the left on both sides of the equation $\mathbf{C}_L^\dagger \psi_b^e(y) =\mathbf{0}$, we get
\begin{equation}
[11.2  \delta^2\mathbbm{1} + ( -3.2 \delta^2+\sqrt{3}\delta \partial_y) \sigma_x] f_b^e(y)\begin{pmatrix}a\\ b\end{pmatrix}=\mathbf{0}.
\end{equation}
Choosing $(a,b) = (1,1)/\sqrt{2}$, the equation becomes a scalar first order differential equation
\begin{equation}
\label{Eq:fbe}
\sqrt{3} \partial_y f_b^e(y) + 8 \delta f_b^e(y) = 0 \Rightarrow f_b^u(y) = \begin{cases}c_{1>}^e e^{-8/\sqrt{3} \int_0^y dy'\delta(y')}, & y>0 \\
c_{1<}^e e^{-8/\sqrt{3} \int_0^y dy'\delta(y')}, & y<0
\end{cases},
\end{equation}
where $c_{1>}^e$ and $c_{1<}^e$ are constants of integration. To find the relation between these two constants, we have to use appropriate boundary condition. The claim is that $c_{1>}^e = c_{1<}^e$. To see this we first recall that we wrote the low energy theory around $\mathbf{q} = (0,2\pi/\sqrt{3})$, as a result the sign of the displacements changes from one unit cell to the next in the direction $(1/2,\sqrt{3}/2)$. Moreover, the low energy theory was written in the displacement basis $\mathbf{U}_L = \{1/\sqrt{3}\{0,-1,0,0,0,1,0,0,1\}^T,1/\sqrt{3}\{1,0,0,0,-1,0,0,-1,0\}^T\}$, and we found that on each side of the domain wall the state of self stress is $(1,1)/\sqrt{2}$ in this basis. With this information, we turn to Fig.~\ref{Fig:figureSb}(b). In unit cell 1 of Fig.~\ref{Fig:figureSb}(b), the tensions/compressions of the bonds according to the basis $\mathbf{U}_L$ are shown. From this, if want to keep the nodes 1-3 force free and want forces only perpendicular to displacement directions for nodes 4-5, the only possible tensions in all other bonds are shown Fig.~\ref{Fig:figureSb}(b). From Fig.~\ref{Fig:figureSb}(b), we see that tensions/compressions are opposite in unit cell 2 and 3. However, that is already taken care of by $\mathbf{q} = (0,2\pi/\sqrt{3})$ in our low energy theory. Therefore, the boundary condition is satisfied by $c_{1>}^e = c_{1<}^e$. Hence
\begin{equation}
    f_b^e(y) = c_{1}^e e^{-8/\sqrt{3}\int_0^y dy'\delta(y')},
\end{equation}
where $c_1^e$ is a constant chosen to normalize zero mode $\psi_b^e(y)$. Therefore, the expression of the state of self stress localized at the domain wall is $\psi_b^e(y) = c_1^e e^{-8/\sqrt{3} \int_0^y dy'\delta(y')}(1,1)^T/\sqrt{2}$. We can check how the zero frequency state of self stress $\psi_b^e(y)$ transforms under the effective mirror operator $\mathbf{M}_e^L$:
\begin{equation}
\mathbf{M}_e^L \psi_b^e(y) = \begin{pmatrix}0& 1\\1 & 0 \end{pmatrix}\frac{f_b^e(y)}{\sqrt{2}}\begin{pmatrix}1\\1 \end{pmatrix} = \frac{f_b^e(y)}{\sqrt{2}}\begin{pmatrix}1\\1 \end{pmatrix} = \psi_b^e(y),
\end{equation}
meaning $\psi_b^e(y)$ is even under mirror $m_x$.

The next question that we can ask is how the frequency of these edge modes would vary from $0$ as go away from $k_x = 0$ perturbatively. We can estimate this easily by projecting the effective ``square root" Hamiltonian in the basis $\{\psi_b^u(y),\psi_b^e(y)\}e^{i k_x x}$ is:
\begin{equation}
\begin{split}
\mathcal{H}_{dw}^b &= \begin{pmatrix}
\int_{-\infty}^\infty dy \psi_b^u(y)^T \begin{pmatrix}
\mathbf{0} & \mathbf{C}_L^\dagger\\
\mathbf{C}_L & \mathbf{0}
\end{pmatrix}\psi_b^u(y) & \int_{-\infty}^\infty dy \psi_b^u(y)^T \begin{pmatrix}
\mathbf{0} & \mathbf{C}_L^\dagger\\
\mathbf{C}_L & \mathbf{0}
\end{pmatrix}\psi_b^e(y)\\
\int_{-\infty}^\infty dy \psi_b^e(y)^T \begin{pmatrix}
\mathbf{0} & \mathbf{C}_L^\dagger\\
\mathbf{C}_L & \mathbf{0}
\end{pmatrix}\psi_b^u(y) & \int_{-\infty}^\infty dy \psi_b^e(y)^T \begin{pmatrix}
\mathbf{0} & \mathbf{C}_L^\dagger\\
\mathbf{C}_L & \mathbf{0}
\end{pmatrix}\psi_b^e(y)
\end{pmatrix} \\
&=  \begin{pmatrix} 0 &  -ik_x \int_{-\infty}^\infty dy f_b^u(y) f_b^e(y)\delta(y)\\ ik_x \int_{-\infty}^\infty dy f_b^u(y) f_b^e(y)\delta(y) & 0\end{pmatrix} =  \begin{pmatrix} 0 &  -iA_bk_x \\ iA_bk_x & 0\end{pmatrix},
\end{split}
\end{equation}
where $A_b = \int_{-\infty}^\infty dy f_b^u(y) f_b^e(y)\delta(y)$. The eigenvalues of this matrix $\mathcal{H}_{dw}^b$ are $\pm|A_bk_x|$, meaning that the edge spectrum is gapless. We can ask if we can add any other term to $\mathcal{H}_{dw}^b$ without breaking the mirror symmetry $m_x$ such that the edge is gapped. The answer is no. To see this, we first note that the representation of the mirror $m_x$ in the basis $\{\psi_b^u(y),\psi_b^e(y)\}$ is:
\begin{equation}
\mathbf{M}_{dw}^b = \begin{pmatrix} -1 & 0\\0 & 1\end{pmatrix} = -\sigma_z.
\end{equation}
Since we are requiring $[\mathcal{H}_{dw}^b,\mathbf{M}_{dw}^b] = 0$ the only term that we can add is proportional to $\mathbbm{1}$ which is not allowed by the chiral symmetry. This essentially means that since the state of self stress is even whereas the zero mode is odd under the mirror, they cannot couple to each other to gap the edge unless the mirror symmetry is broken.

\underline{Case 2: $\text{sgn}(\delta) = -\text{sgn}(y)$}: This is the case at the top domain wall in Fig.~\sss{3} of the main text. To obtain the zero mode, we choose the same form of the zero mode $\psi_t^u(y) =f_t^u(y)(a,b)^T$, and following the same steps ad in Case 1 get to the equation:
\begin{equation}
[11.2  \delta^2\mathbbm{1} + ( 3.2 \delta^2+\sqrt{3}\delta \partial_y) \sigma_x] f_t^u(y)\begin{pmatrix}a\\ b\end{pmatrix}=\mathbf{0}.
\end{equation}
However, this time we choose $(a,b) = (1,-1)/\sqrt{2}$. Consequently, the equation becomes a scalar first order differential equation
\begin{equation}
\sqrt{3} \partial_y f_t^u(y) - 8 \delta f_t^u(y) = 0 \Rightarrow f_t^u(y) = c_2^u\text{sgn}(y) e^{8/\sqrt{3} \int_0^y dy'\delta(y')},
\end{equation}
where the factor $\text{sgn}(y)$ is due to similar boundary condition as in case 1. Here $c_2^u$ is a constant chosen to normalize zero mode $\psi_t^u(y)$. The function $f_t^u(y)$ is exponentially decaying away from $y = 0$ due to the fact that $\delta<0$ for $y>0$ and $\delta>0$ for $y<0$. Therefore, zero frequency edge mode of $\mathbf{C}_L(k_x = 0)$ is $\psi_t^u(y) = c_2^u \text{sgn}(y) e^{8/\sqrt{3} \int_0^y dy'\delta(y')}(1,-1)^T/\sqrt{2}$. We can check how the zero frequency edge mode $\psi_t^u(y)$ transforms under the effective mirror operator $\mathbf{M}_u^L$:
\begin{equation}
\mathbf{M}_u^L \psi_t^u(y) = \begin{pmatrix}0& -1\\-1 & 0 \end{pmatrix}\frac{f_t^u(y)}{\sqrt{2}}\begin{pmatrix}1\\-1 \end{pmatrix} = \frac{f_t^u(y)}{\sqrt{2}}\begin{pmatrix}1\\-1 \end{pmatrix} = \psi_t^u(y),
\end{equation}
meaning $\psi_t^u(y)$ is even under mirror $m_x$. This matches with the plot in Fig.~\sss{3}(b).

Next we find the state of self stress at this domain wall. For that we will work with the matrix $\mathbf{C}_L^\dagger$:
\begin{equation}
\mathbf{C}_L^\dagger = 11.2  \delta^2 \sigma_z - i\delta k_x \sigma_x + ( -3.2 i \delta^2-\sqrt{3}\delta k_y) \sigma_y \rightarrow 11.2  \delta^2 \sigma_z - i\delta k_x \sigma_x + ( -3.2 i \delta^2+i\sqrt{3}\delta\partial_y) \sigma_y
\end{equation}
Similar to before, we are going to consider solution of the form $\psi_t^e(y) =f_t^e(y)(a,b)^T$ with exponentially localized $f_t^e(y)$. Setting $k_x = 0$ and multiplying by $\sigma_z$ from the left on both sides of the equation $\mathbf{C}_L^\dagger \psi_t^e(y) =\mathbf{0}$, we get
\begin{equation}
[11.2  \delta^2\mathbbm{1} + ( -3.2 \delta^2+\sqrt{3}\delta \partial_y) \sigma_x] f_t^e(y)\begin{pmatrix}a\\ b\end{pmatrix}=\mathbf{0}.
\end{equation}
Choosing $(a,b) = (1,-1)/\sqrt{2}$, the equation becomes a scalar first order differential equation
\begin{equation}
\sqrt{3} \partial_y f_t^e(y) - 14.4 \delta f_t^e(y) = 0 \Rightarrow f_t^u(y) = c_1^e e^{14.4/\sqrt{3} \int_0^y dy'\delta(y')},
\end{equation}
where $c_2^e$ is a constant chosen to normalize zero mode $\psi_t^e(y)$. Therefore, the expression of the state of self stress localized at the domain wall is $\psi_t^e(y) = c_2^e e^{14.4/\sqrt{3} \int_0^y dy'\delta(y')}(1,-1)^T/\sqrt{2}$. We can check how the zero frequency state of self stress $\psi_b^e(y)$ transforms under the effective mirror operator $\mathbf{M}_e^L$:
\begin{equation}
\mathbf{M}_e^L \psi_t^e(y) = \begin{pmatrix}0& 1\\1 & 0 \end{pmatrix}\frac{f_t^e(y)}{\sqrt{2}}\begin{pmatrix}1\\-1 \end{pmatrix} = -\frac{f_t^e(y)}{\sqrt{2}}\begin{pmatrix}1\\1 \end{pmatrix} = -\psi_t^e(y),
\end{equation}
meaning $\psi_t^e(y)$ is odd under mirror $m_x$.

The next question that we can ask is how the frequency of these edge modes would vary from $0$ as go away from $k_x = 0$ perturbatively. We can estimate this easily by projecting the effective ``square root" Hamiltonian in the basis $\{\psi_t^u(y),\psi_t^e(y)\}e^{ik_x x}$ is:
\begin{equation}
\begin{split}
\mathcal{H}_{dw}^t &= \begin{pmatrix}
\int_{-\infty}^\infty dy \psi_t^u(y)^T \begin{pmatrix}
\mathbf{0} & \mathbf{C}_L^\dagger\\
\mathbf{C}_L & \mathbf{0}
\end{pmatrix}\psi_t^u(y) & \int_{-\infty}^\infty dy \psi_t^u(y)^T \begin{pmatrix}
\mathbf{0} & \mathbf{C}_L^\dagger\\
\mathbf{C}_L & \mathbf{0}
\end{pmatrix}\psi_t^e(y)\\
\int_{-\infty}^\infty dy \psi_t^e(y)^T \begin{pmatrix}
\mathbf{0} & \mathbf{C}_L^\dagger\\
\mathbf{C}_L & \mathbf{0}
\end{pmatrix}\psi_t^u(y) & \int_{-\infty}^\infty dy \psi_t^e(y)^T \begin{pmatrix}
\mathbf{0} & \mathbf{C}_L^\dagger\\
\mathbf{C}_L & \mathbf{0}
\end{pmatrix}\psi_t^e(y)
\end{pmatrix} \\
&=  \begin{pmatrix} 0 &  -ik_x \int_{-\infty}^\infty dy f_t^u(y) f_t^e(y)\delta(y)\\ ik_x \int_{-\infty}^\infty dy f_t^u(y) f_t^e(y)\delta(y) & 0\end{pmatrix} =  \begin{pmatrix} 0 &  -iA_tk_x \\ iA_tk_x & 0\end{pmatrix},
\end{split}
\end{equation}
where $A_t = \int_{-\infty}^\infty dy f_t^u(y) f_t^e(y)\delta(y)$. The eigenvalues of this matrix $\mathcal{H}_{dw}^t$ are $\pm|A_tk_x|$, meaning that the edge spectrum is gapless. We can ask if we can add any other term to $\mathcal{H}_{dw}^t$ without breaking the mirror symmetry $m_x$ such that the edge is gapped. The answer is no. To see this, we first note that the representation of the mirror $m_x$ in the basis $\{\psi_t^u(y),\psi_t^e(y)\}$ is:
\begin{equation}
\mathbf{M}_{dw}^t = \begin{pmatrix} 1 & 0\\0 & -1\end{pmatrix} = \sigma_z.
\end{equation}
Since we are requiring $[\mathcal{H}_{dw}^t,\mathbf{M}_{dw}^t] = 0$ the only term that we can add is proportional to $\mathbbm{1}$ which is not allowed by the chiral symmetry. This essentially means that since the state of self stress is even whereas the zero mode is odd under the mirror, they cannot couple to each other to gap the edge unless the mirror symmetry is broken.


\begin{figure}[b]
\includegraphics{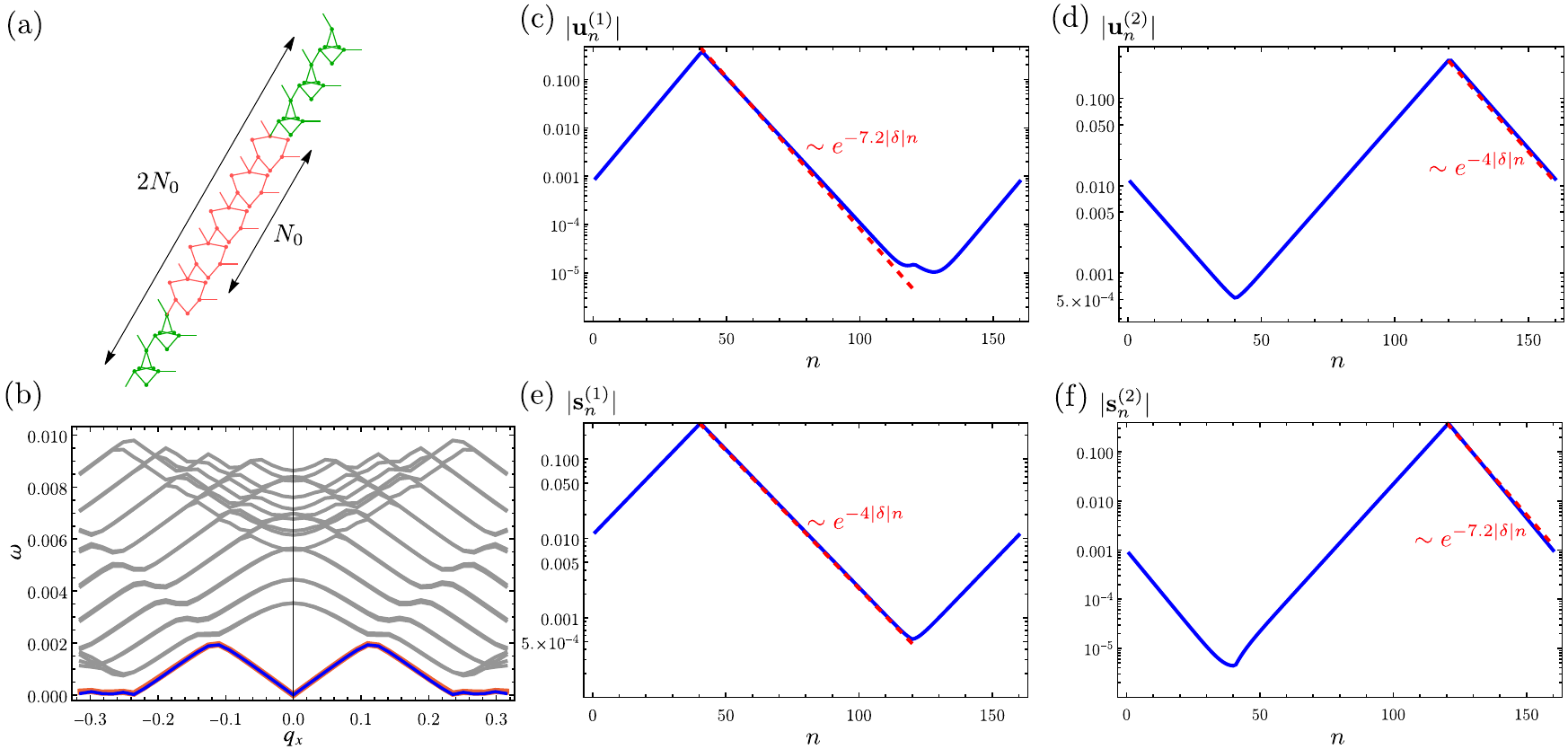}
\caption{Comparison between numerical calculation and analytical low energy theory prediction for the decay rate away from the domain wall of the zero modes and states of self stress. (a) Schematic of the system used for numerical calculation. The red region has $\delta>0$ whereas the green region has $\delta<0$. There are $N_0$ unit cells in each region. All numerical calculations in (b-f) are done for $N_0=80$ and $\delta = 1/50$ in red region and $\delta = -1/50$ in green region. Periodic boundary condition is employed in direction $(1/2,\sqrt{3}/2)$, whereas Bloch-periodic boundary condition $\mathbf{u}(\mathbf{x}+(1,0)) =\mathbf{u}(\mathbf{x}) e^{iq_x}$ is employed in $(1,0)$ direction. (b) shows the band eigenfrequencies $\omega$ as a function of $q_x$. Plotted in grey are the bulk modes, whereas the modes corresponding to the blue and the red bands are concentrated at the bottom and the top domain walls. In (c-d) the norms of the zero modes ($|\mathbf{u}^{(i)}_{n}|$ as defined in Eq.~\eqref{Eq:ZMNorm}) are plotted in blue as a function of unit cell number. The zero mode in (c) is concentrated at the bottom domain wall whereas the zero mode in (d) is concentrated at the top domain wall. In (e-f) the norms of the states of self stress ($|\mathbf{s}^{(i)}_{n}|$ as defined in Eq.~\eqref{Eq:SSSNorm}) are plotted in blue as a function of unit cell number. The state of self stress in (e) is concentrated at the bottom domain wall whereas the state of self stress in (f) is concentrated at the top domain wall. In each of (c-f) the theoretical decay rate is plotted in red dashed line with corresponding exponential factor written in red beside it.}
\label{Fig:figureS1}
\end{figure}

We validate the results from low energy theory described above with numerical results in Fig.~\ref{Fig:figureS1}. For numerical calculation, we created a system just like Fig.~\sss{3} of main text. The low energy theory works for small $\delta$, but for small $\delta$, the system is not fully gapped at $\omega=0$ as shown in Fig.~\sss{2} of main text. However, fortunately, the system is gapped near $q_x = 0$ for small $\delta$ which is region where the low energy theory works anyway. Anticipating that for small $\delta$ the zero frequency edge modes and states of self stress will decay slowly away from the domain wall, we took a system with $N_0 = 80$ unit cells in each phase ($N_0$ is defined in Fig.~\ref{Fig:figureS1}(a)). The boundary condition is chosen to be the same as in Fig.~\sss{3} of main text. Two zero modes appear at $q_x = 0$ as shown in Fig.~\ref{Fig:figureS1}(b). In Fig.~\ref{Fig:figureS1}(c) and (d), we plot in blue solid lines the norm of the two zero modes $\mathbf{u}^{(1)}$ and $\mathbf{u}^{(2)}$ in each unit cell as a function of unit cell number, where the norm of each zero mode in $n$th unit cell is defined as
\begin{equation}
\label{Eq:ZMNorm}
    |\mathbf{u}^{(i)}_{n}| = \sqrt{\sum_{j=1}^{9} \left(\mathbf{u}^{(i)}_{n,j}\right)^2},
\end{equation}
where the sum goes over the $9$ degrees of freedom per unit cell. The unit cells are enumerated from bottom towards top, i.e., the unit cell number 1 at the bottom most one and the unit cell number 160 is the top most one. The red dashed lines show the exponential decay predicted by low energy theory. Note that the exponential factors from the low energy theory were $e^{-14.4 |\delta|y/\sqrt{3}}$ and $e^{-8 |\delta|y/\sqrt{3}}$. To plot it as function of unit cell, we recognize that each unit cell is of length $\sqrt{3}/2$. Therefore, as function of unit cell number $n$ these factors become $e^{-7.2 |\delta|n}$ and $e^{-4 |\delta|n}$. The decay rates of the zero modes from the numerical calculation match very well with the theoretical prediction. Similarly, In Fig.~\ref{Fig:figureS1}(e) and (f), we plot in blue solid lines the norm of the two states of self stress $\mathbf{s}^{(1)}$ and $\mathbf{s}^{(2)}$ in each unit cell as a function of unit cell number, where the norm of each state of self stress in $n$th unit cell is defined as
\begin{equation}
\label{Eq:SSSNorm}
    |\mathbf{s}^{(i)}_{n}| = \sqrt{\sum_{j=1}^{9} \left(\mathbf{s}^{(i)}_{n,j}\right)^2},
\end{equation}
where the sum goes over the $9$ degrees of freedom per unit cell. The red dashed lines show the exponential decay predicted by low energy theory. Again, the decay rates of the zero modes from the numerical calculation match very well with the theoretical prediction.

\section{Corner states from the low energy theory}\label{Sec:LowEnergyCorner}
\begin{figure}[b]
\includegraphics{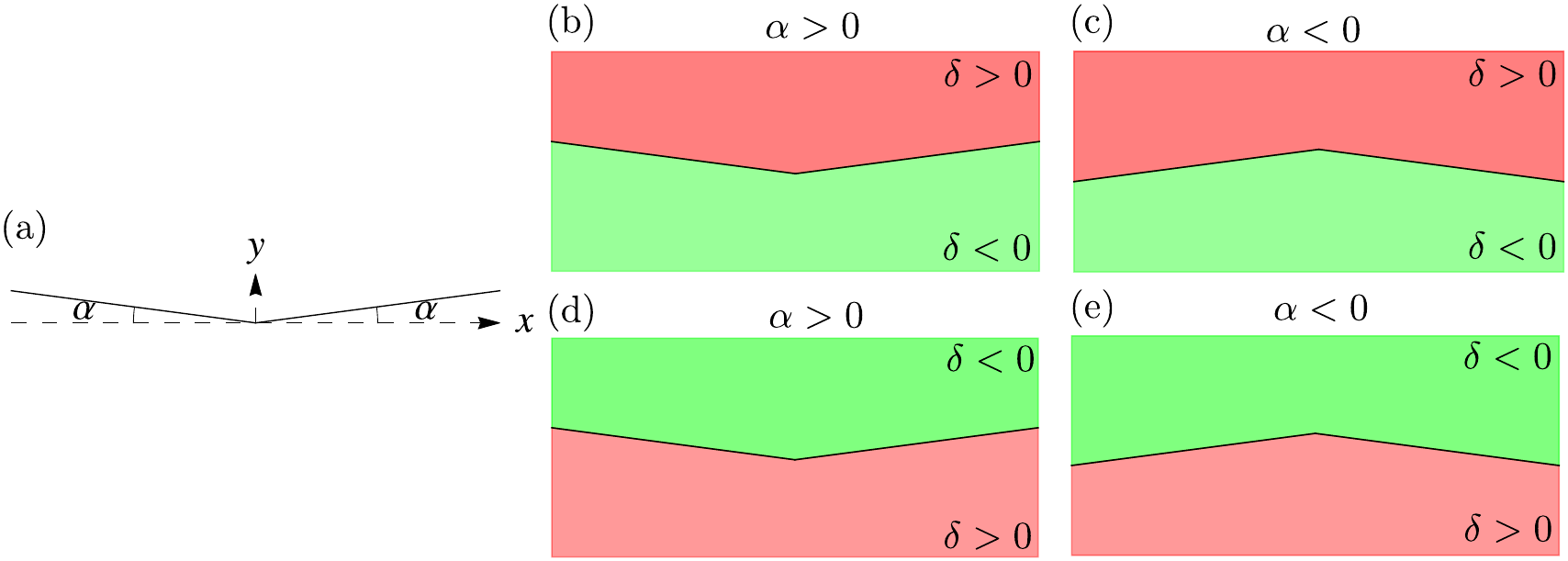}
\caption{Corners. (a) Creating a corner from a straight domain wall $ y = 0$ by tilting $x>0$ and $x<0$ sections in the opposite direction by angle $\alpha$. (b)-(e) show four different cases depending on the value of $\alpha$ as well as the which phase is above or below the domain wall. The red and green color denote phases with $\delta>0$ and $\delta<0$ respectively.} 
\label{Fig:figureS2}
\end{figure}
Following the analysis in~\citesupp{schindler2018higher1,neupert2018topological1}, now we tilt the $x>0$ and $x<0$ sections of the domain wall $y = 0$ in the opposite direction by angle $\alpha$ and $-\alpha$ respectively to break the mirror symmetry on the domain walls but keep the mirror symmetry at the corner at $x = 0 = y$ as shown in Fig.~\ref{Fig:figureS2}(a). Now, there will be the four cases shown in Fig.~\ref{Fig:figureS2}(b)-(e). We will consider Fig.~\ref{Fig:figureS2}(b) and (c) together first, and then discuss cases (d) and (e). Before considering each of the cases in detail, let us discuss the effect of breaking mirror symmetry far away from the corner $x= 0 =y$. To keep the problem analytically tractable, we will consider $\alpha \ll 1$ and the domain wall is still very close being parallel to the $x$-axis. We will start from the domain wall ``square root" Hamiltonian $\mathcal{H}_{dw}$ and replace $ik_x$ with $\partial_x$ in $\mathcal{H}_{dw}$ since we are breaking translation symmetry in the $x$-direction by creating the corner. More importantly, we can now add extra terms to $\mathcal{H}_{dw}$ since we have broken the mirror symmetry on the domain walls. The only nontrivial term that break mirror symmetry ($\sigma_z \mathcal{H}_{dw}(k_x) \sigma_z = \mathcal{H}_{dw}(-k_x)$ since $\sigma_z$ is the mirror operator in the space of the domain wall modes) while maintaining time reversal symmetry ($\mathcal{H}_{dw}^*(k_x) = \mathcal{H}_{dw}(-k_x)$) is $m \sigma_x$. This mass $m$ has to be proportional to $\alpha$ to the lowest order in $\alpha$ since when $\alpha = 0$, the mirror symmetry is restored. The mass $m$ gaps domain wall spectrum at $\omega = 0$. However, since the system is still mirror symmetric about $x = 0$, $m(x) = -m(-x)$ such that we have mirror symmetry about $x=0$: $\sigma_z m(x)\sigma_x \sigma_z = -m(x)\sigma_x = m(-x)\sigma_x$. Therefore, the modified Hamiltonian $\mathcal{H}_{c}$ for the corner is
\begin{equation}
\label{Eq:CornerH}
\mathcal{H}_{c} = -iA \sigma_y\partial_x + m(x)\sigma_x,
\end{equation}
where $A$ takes value $A_b$ for the cases in Fig.~\ref{Fig:figureS2}(b-c) and $A_t$ for the cases in Fig.~\ref{Fig:figureS2}(d-e). This is readily recognizable as the low energy theory of the Su-Schriffer-Heager (SSH) model~\citesupp{jackiw2007fractional1}.

\underline{Cases in Fig.~\ref{Fig:figureS2}(b-c)}: These two are obtained from case 1 in the previous section by deforming the domain in opposite direction.Therefore, the corner Hamiltonian $\mathcal{H}_{c}$ in these two cases are obtained by modifying $\mathcal{H}_{dw}^b$ (which is written in the basis $\{\psi_b^u, \psi_b^e\}$). In (b), the slope of the domain wall is positive (negative) when $x>0$ ($x<0$). The configuration in (c) is opposite, i.e., the slope of the domain wall is positive (negative) when $x<0$ ($x>0$). As a result, $m^{(b)}(x) = -m^{(c)}(x)$. Note that Fig.~\ref{Fig:figureS2}(b) is situation at the bottom corner of Fig.~\ref{Fig:figure4}(c-d) in the main text, whereas Fig.~\ref{Fig:figureS2}(c) corresponds to the top corner of Fig.~\ref{Fig:figure4}(a-b). We seek solutions of the equation $\mathcal{H}_{c}\tilde{\psi}_b(x) =  \begin{pmatrix}
0 & -A_b\partial_x +m(x)\\A_b\partial_x +m(x) & 0
\end{pmatrix}\tilde{\psi}_b(x) = \mathbf{0}$ of the form $\tilde{\psi}_b^u(x) = g_b^u(x)(1,0)^T$ and $\tilde{\psi}_b^e(x) = g_b^e(x)(0,1)^T$. The first one would be a zero mode, and second one would be a state of self stress. Plugging these, we get
\begin{equation}
\begin{split}
A_b \partial_x g_b^u(x)+m(x)g_b^u(x) &= 0 \Rightarrow g_b^u(x) = a_b^u e^{-\int_0^x dx' m(x')/A_b},\\
-A_b \partial_x g_b^e(x)+m(x)g_b^e(x) &= 0 \Rightarrow g_b^e(x) = a_b^e e^{\int_0^x dx' m(x')/A_b}.\\
\end{split}
\end{equation}
Note that the full solution for the zero mode (state of self stress) is then $\psi_c^u(x,y)= g_b^u(x) \psi_b^u(y)  = g_b^u(x) f_b^u(y)(1,1)^T/\sqrt{2}$ ($\psi_c^e(x,y) = g_b^e(x) f_b^e(y)(1,1)^T/\sqrt{2}$). A few points are in order here. First, $g_b^u(x)$ is exponentially decay away from $x = 0$ if $m(x)/A_b>0 \text{ and } m(x)/A_b<0$ for $x>0 \text{ and } x<0$ respectively, whereas it grows exponentially away from $x = 0$ if $m(x)/A_b<0 \text{ and } m(x)/A_b>0$ for $x>0 \text{ and } x<0$ respectively. One of them is the case for Fig.~\ref{Fig:figureS2}(b), the other for Fig.~\ref{Fig:figureS2}(c). Therefore, if in one of these subfigures, there is a zero mode exponentially decaying away from $x = 0$, there would be a zero mode exponentially growing away from $x = 0$. In the case, where the zero mode is exponentially grows away from $x = 0$, it will be exponentially localized at the other ends of the domain wall. This is exactly why in case of \sss{Fig.~4(b)} the corner mode is localized at top corner, whereas in \sss{Fig.~4(d)} the corner mode is localized at the right and left corners (these are the other two ends of the domain walls). Moreover, in both cases the zero mode is odd under mirror $m_x$ passing through the top corner since the basis function for the zero mode $\psi_b^u(y) = f_b^u(y)(1,1)^T/\sqrt{2}$ is odd under $\mathbf{M}_u^L$.

\underline{Cases in Fig.~\ref{Fig:figureS2}(d-e)}: These two are obtained from case 2 in the previous section by deforming the domain in opposite direction. Therefore, the corner Hamiltonian $\mathcal{H}_{c}$ in these two cases are obtained by modifying $\mathcal{H}_{dw}^t$ (which is written in the basis $\{\psi_t^u, \psi_t^e\}$).  In (d), the slope of the domain wall is positive (negative) when $x>0$ ($x<0$). The configuration in (e) is opposite, i.e., the slope of the domain wall is positive (negative) when $x<0$ ($x>0$). As a result, $m^{(d)}(x) = -m^{(e)}(x)$. Note that Fig.~\ref{Fig:figureS2}(d) is situation at the bottom corner of Fig.~\ref{Fig:figure4}(a-b) in the main text, whereas Fig.~\ref{Fig:figureS2}(e) corresponds to the top corner of Fig.~\ref{Fig:figure4}(c-d) in the main text. We seek solutions of the equation $\mathcal{H}_{c}\tilde{\psi}_t(x) =  \begin{pmatrix}
0 & -A_t\partial_x +m(x)\\A_t\partial_x +m(x) & 0
\end{pmatrix}\tilde{\psi}_t(x) = \mathbf{0}$ of the form $\tilde{\psi}_t^u(x) = g_t^u(x)(1,0)^T$ and $\tilde{\psi}_t^e(x) = g_t^e(x)(0,1)^T$. The first one would be a zero mode, and second one would be a state of self stress. Plugging these, we get
\begin{equation}
\begin{split}
A_t \partial_x g_t^u(x)+m(x)g_t^u(x) &= 0 \Rightarrow g_t^u(x) = a_t^u e^{-\int_0^x dx' m(x')/A_t},\\
-A_t \partial_x g_t^e(x)+m(x)g_t^e(x) &= 0 \Rightarrow g_t^e(x) = a_t^e e^{\int_0^x dx' m(x')/A_t}.\\
\end{split}
\end{equation}
Note that the full solution for the zero mode (state of self stress) is then $psi_c^u(x,y)= g_t^u(x) \psi_t^u(y)  = g_t^u(x) f_t^u(y)(1,-1)^T/\sqrt{2}$ ($\psi_c^e(x,y) = g_t^e(x) f_t^e(y)(1,-1)^T/\sqrt{2}$). A few points are in order here. First, $g_t^u(x)$ is exponentially decay away from $x = 0$ if $m(x)/A_t>0 \text{ and } m(x)/A_t<0$ for $x>0 \text{ and } x<0$ respectively, whereas it grows exponentially away from $x = 0$ if $m(x)/A_t<0 \text{ and } m(x)/A_t>0$ for $x>0 \text{ and } x<0$ respectively. One of them is the case for Fig.~\ref{Fig:figureS2}(d), the other for Fig.~\ref{Fig:figureS2}(e). Therefore, if in one of these subfigures, there is a zero mode exponentially decaying away from $x = 0$, there would be a zero mode exponentially growing away from $x = 0$. In the case, where the zero mode is exponentially grows away from $x = 0$, it will be exponentially localized at the other ends of the domain wall. This is exactly why in case of \sss{Fig.~4(a)} the corner mode is localized at bottom corner, whereas in \sss{Fig.~4(c)} the corner mode is localized at the right and left corners (these are the other two ends of the domain walls). Moreover, in both cases the zero mode is even under mirror $m_x$ passing through the top corner since the basis function for the zero mode $\psi_t^u(y) = f_t^u(y)(1,-1)^T/\sqrt{2}$ is even under $\mathbf{M}_u^L$. 

Similar calculations can be done to obtain states of self stress localized at corners.

\section{Corner modes where the mirror symmetry is broken at the corner}\label{Sec:CornerBrokenMirror}
\begin{figure*}[h]
\includegraphics{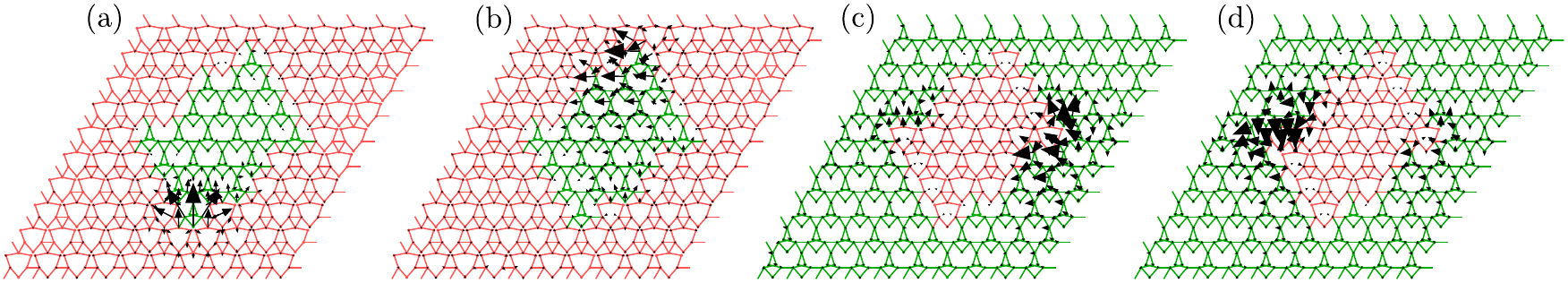}
\caption{Corner modes in systems with diamond shaped island of one phase inside the other phase.. In each panel, the part of the system in red (green) has $\delta = 1/3$ ($\delta = -13/42$). The angle of tilt of the domain walls on the left and right of the corners are not the same unlike Fig.~\sss{4} of main text. We applied periodic boundary conditions in all cases. The black arrows show the displacement field corresponding to the zero modes. The zero modes are still localized at the corners just like they were in Fig.~\sss{4} of main text, however they are mirror symmetric since the mirror at the corners is broken due to different angle of tilt on each side of the corners.
} 
\label{Fig:figureCornerBrokenMirror}
\end{figure*}
\bibliographystylesupp{apsrev4-1}
\bibliographysupp{refer.bib}
\end{document}